\documentclass{article}

\usepackage{arxiv}

\usepackage[utf8]{inputenc} 
\usepackage[T1]{fontenc}    
\usepackage{hyperref}       
\usepackage{url}            
\usepackage{booktabs}       
\usepackage{amsfonts}       
\usepackage{nicefrac}       
\usepackage{microtype}      
\usepackage{lipsum}
\usepackage{graphicx}
\graphicspath{ {./images/} }

\usepackage{amsmath}
\usepackage{longtable,tabularx}
\usepackage{float}
\usepackage{multirow}
\usepackage{caption}
\usepackage{gensymb}
\usepackage{bm}
\usepackage{tabularx,booktabs}
\newcolumntype{Y}{>{\centering\arraybackslash}X}
\newcolumntype{P}[1]{>{\centering\arraybackslash}p{#1}}
\usepackage[section]{placeins}
\usepackage{setspace}
\usepackage{soul}
\usepackage{natbib}
\usepackage{xcolor}

\title{High-speed micro-actuation in a supersonic dual-stream jet flow}

\author{
 Melissa Yeung \\
  Mechanical and Aerospace Engineering\\
  Syracuse University\\
  Syracuse, NY 13244 \\
  \texttt{meyeung@syr.edu} \\
   \And
 Datta V. Gaitonde \\
  Mechanical and Aerospace Engineering\\
  The Ohio State University\\
  Columbus, OH 43210 \\
  \texttt{gaitonde.3@osu.edu} \\
  \And
 Yiyang Sun \\
  Mechanical and Aerospace Engineering\\
  Syracuse University\\
  Syracuse, NY 13244 \\
  \texttt{ysun58@syr.edu} \\
}

\begin{document}
\maketitle
\begin{abstract}
Supersonic shear layers experience instabilities that generate significant adverse effects; in complex configurations, these instabilities have global impacts as they foster compounding complications with other independent flow features. 
We consider the flow near the exit of a dual-stream rectangular nozzle, in which the supersonic core and sonic bypass streams mix downstream of a splitter plate trailing edge (SPTE) just above an adjacent deck representative of a wing surface. 
Active flow control is explored to alleviate the prominent tone associated with vortices shed at the SPTE; these vortices also initiate an unsteady shock system that affects the entire flow field through a shock-induced separation and the downstream evolution of plume shear layers. 
Resolvent analysis is performed on the baseline flow. The identified optimal location guides the placement of steady-blowing micro-jet actuators. A Navier--Stokes-based parametric study is carried out to consider various actuation angles and locations. Since the resolvent analysis fundamentally investigates the input-output dynamics of a system, it is also utilized to uncover actuation-induced changes in the forcing-response dynamics.
Spectral analysis shows that the baseline flow fluctuating energy is concentrated in the shedding instability. Actuating at optimal angles based on location disperses this energy into various flow features; this affects the shedding itself, and the structure and unsteadiness of the shock system and thus the response of the deck and nozzle wall boundary layers and the plume. The resolvent analysis indicates, and Navier-Stokes solutions confirm, that favorable control is obtained by either indirectly or directly mitigating the baseline instability.
\end{abstract}


\section{Introduction}
\label{sec:intro}

\begin{figure}
    \centering
    \includegraphics[width=0.75\textwidth]{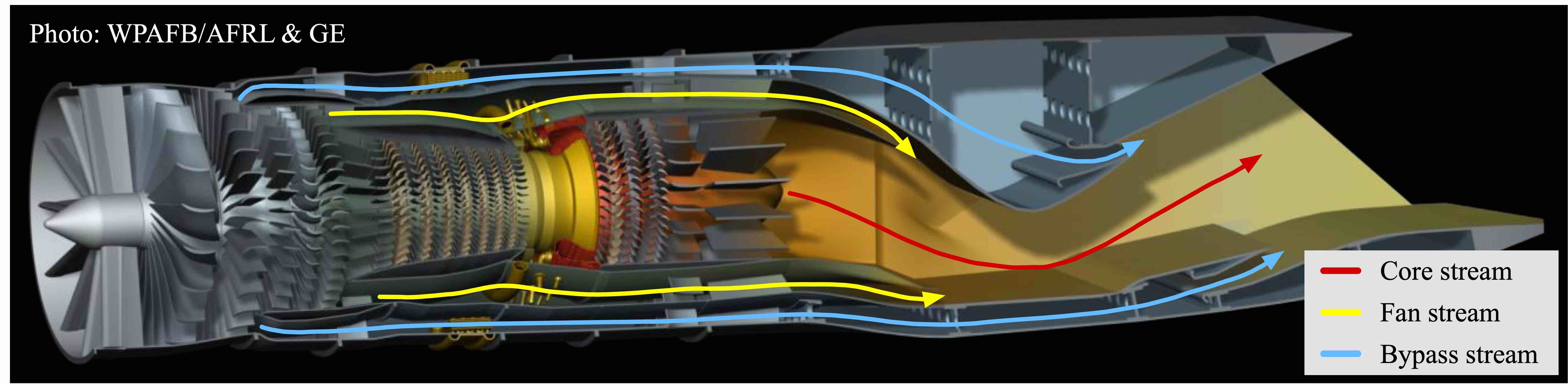}
    \caption{Illustration of generic three-stream engine (\cite{simmons2009design}).}
    \label{fig:engine}
\end{figure}

Proposed next-generation supersonic aircraft consider progressively complex engine configurations to achieve demanding flight requirements, motivating the need to understand and control ever more intricate jet flow physics. A modern variable-cycle engine design, of increasing interest over the last decade, is discussed by \cite{simmons2009design} and shown schematically in figure~\ref{fig:engine}. In addition to a rectangular cross-section to facilitate easier airframe integration and drag reduction, the arrangement utilizes a single-sided expansion ramp nozzle (SERN) with a unique independently modulated, cooled bypass stream to continuously match inlet airflow and engine demand while simultaneously serving as a cool heat sink to dissipate aircraft heat loads. Under suitable conditions, the addition of such streams has also been shown to reduce noise \citep{papamoschou2001directional, berry2016acoustic}. Numerous variations of the multi-stream nozzle design have been previously studied by \cite{hromisin2019}, \cite{papamoschou2016quiet}, \cite{hromisin2019aeroacoustic}, \cite{tinney2018aeroacoustics}, and \cite{ruscher2016noisy} due to the advantages provided by the tertiary bypass stream and promising noise reductions, making this a highly coveted design for implementation into future supersonic air vehicles.

\begin{figure}
    \centering
    \includegraphics[width=1\textwidth]{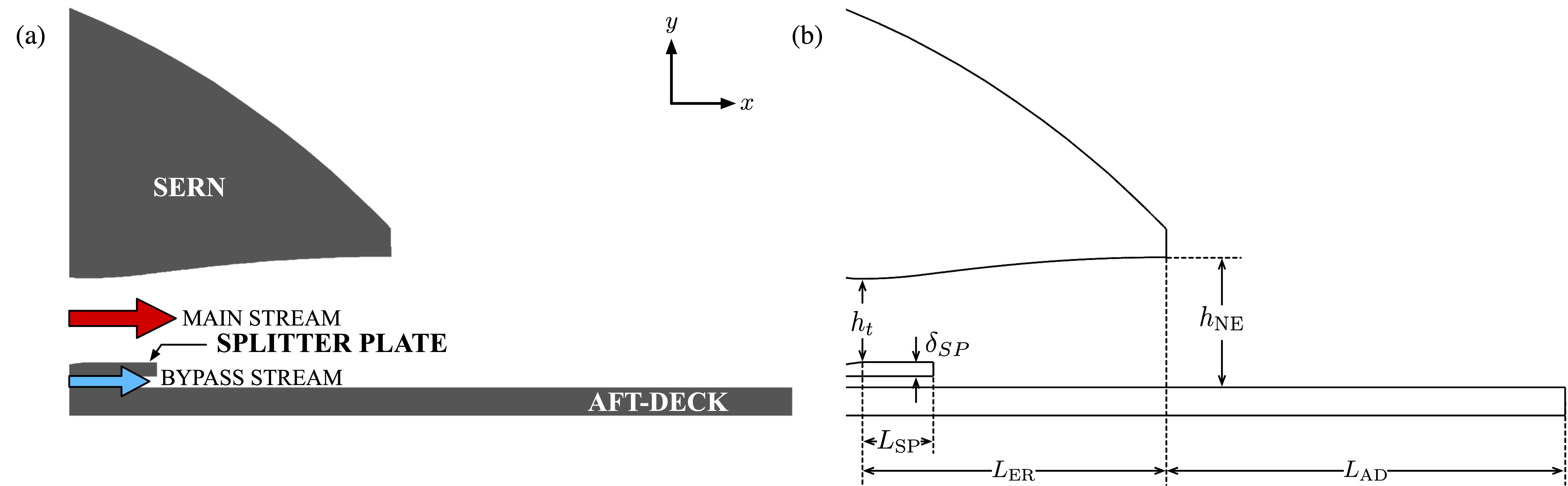}
    \caption{(a) Nozzle configuration and (b) nomenclature}
    \label{fig:nozzle}
\end{figure}

The present study analyzes the supersonic flow produced at the exit of the engine shown in figure \ref{fig:engine}, where it is assumed that the indicated core and fan streams are fully mixed before exiting the nozzle. A cutaway of the nozzle cross-section region of interest is shown in figure \ref{fig:nozzle}(a). Thus, the nozzle contains two canonical flows: a main core stream and a bypass stream with a splitter plate separating the two, while an aft-deck is used to represent the structure of an aircraft wing or fuselage. The cross-section becomes complete with the addition of the upper SERN nozzle surface.

The aerodynamic flow at the engine exit contains various turbulent mixing layers, which have been previously studied by \cite{delville1999examination} and \cite{ukeiley2001examination}. 
Supersonic jets have been known to produce jet noise that is generally associated with turbulent mixing, broadband shock-associated noise (BBSAN), and screech tones. Investigation into jet noise was first pioneered by \cite{lighthill1952sound,lighthill1954sound} and later extended to supersonic turbulent shear layers by \cite{phillips1960generation}, who predicted the sound field radiates as eddy Mach waves that are produced along the shear zone.
Moreover, interactions between the shock cells and shear layer instabilities downstream of the nozzle exit generate acoustic waves that are mostly broadband. In the scenario where the upstream propagating sound waves excite the shear layer near the nozzle lip, screech tones are produced from the closure of a feedback loop as \cite{powell1953mechanism} first suggested. Under certain conditions, harmonics of the screech tone can also be observed (\cite{tam2014harmonics}).

In the present work, the nozzle houses a supersonic shear layer that is generated from the two incoming streams. The flow downstream of the relatively thick splitter plate is characterized by undesirable flow features whose influence permeates other regions in the entire cross-section of the nozzle exit and also the plume, in the form of tonal noise and unsteady surface loading. The main flow features are vortex shedding and its consequences, including the formation of a relatively pronounced unsteady shock system that traverses the flow, interacts with and reflects from the SERN back on the deck surface. Extensive computational and experimental work has been performed in an attempt to control the flow. Because the dominant flow physics of shedding vortices from the splitter plate contributes most to the far-field resonant tone, these efforts have been directed towards perturbing the flow near the splitter plate trailing edge (SPTE). Passive control in the form of a spanwise wavy splitter plate was investigated experimentally by \cite{gist2022exploring} and computationally by \cite{doshi2022passive}. A key finding was that the stream-wise vorticity induced by the wavy splitter plate broke down the coherent structures associated with the signature high-frequency tone found in far-field acoustics, resulting in a reduction of the dominant tone. 

The success of this passive control technique prompted investigation into active control strategies such that similar control outcomes could be achieved, allowing for the potential development of an efficient and flexible sensor-actuator control system capable of operating under various flight conditions and with minimal energy input. Recently, \cite{kelly2023active} implemented micro-jet arrays that blow steady, sonic air into the flow. Because of experimental limitations, the actuators were placed on the aft-deck, beneath the splitter plate trailing edge, rather than on the potentially more effective locations, such as the splitter plate itself. Results with blowing at different angles indicated encouraging control authority on the shock system, which only modestly altered the tone. The present work lifts experimental constraints by performing a parametric study examining how actuator placement at different locations on the splitter plate affects the resonant tone and other flow characteristics.

Unsteady simulations using the Navier-Stokes equations are performed to obtain the data for the baseline and controlled conditions. 
An understanding of the simulated flow fields may be derived from modal analyses of the unsteady data.
Various data-driven modal analysis techniques are appealing for their ability to decompose a turbulent flow field into modes that allow one to identify physically important flow features. \cite{taira2017modal}, \cite{taira2020modal}, and \cite{rowley2017model} have presented an overview and framework for several well-known modal decomposition methods and their applications. 
Proper orthogonal decomposition (POD) is a popular technique first introduced by \cite{lumley1967structure} as a means to extract coherent structures from turbulent flow fields. It seeks an optimal set of basis functions that represent the given dataset, where each POD mode captures a portion of the total energy. In the frequency-based form of POD, known as the spectral proper orthogonal decomposition (SPOD) (\cite{glauser1987coherent}), dominant structures are instead extracted across a range of frequencies. It has proven to be a powerful tool in characterizing a turbulent flow and has seen increasing use across the fluid dynamics community.
The properties of SPOD and its relationship to other approaches have been elaborated by \cite{towne2018spectral}. 
Examples, where SPOD variants have been leveraged to analyze both experimental and numerical datasets, include turbulent jets (\cite{glauser1987coherent, schmidt2018spectral, akamine2022interpretation}), turbulent wakes (\cite{araya2017transition, nidhan2020spectral}), and flows past airfoils (\cite{ribeiro2017identification, abreu2017coherent}).
In the present study, the SPOD is applied to numerical datasets obtained from direct numerical simulations to identify flow structures at each frequency.

To inform actuator placement from a sensitivity and receptivity perspective and to characterize the effects of control, the simulated results are subjected to the resolvent analysis, which yields forcing and response dynamics of the simulated flow. The resolvent analysis is an operator-based modal technique to extract the input-output dynamics of a linear system. It may be viewed as complementary to global linear stability analysis which predicts the asymptotic evolution of disturbances in a base flow through an eigenvalue problem (\cite{mckeon2010critical, taira2017modal}). In contrast, the resolvent analysis predicts forcing and response modes for chosen wavenumbers and frequencies through the singular value decomposition of the resolvent operator, which gives optimal energy amplifications between forcing and response modes. 
This yields physical insights into control designs, and is particularly beneficial when searching for effective active control configurations because the parameter space is extensive and costly to test with resolved simulations.
Moreover, this approach facilitates  investigation of the input-output dynamics of a system under the influence of control.

The paper is organized as follows. The active control configuration, computational approach, and methods used to analyze the supersonic flow generated from a single-sided expansion ramp are detailed in Section \ref{sec:methodSetup}. Results are presented in Section \ref{sec:results}, where instantaneous and mean flow modifications are discussed along with the overall instabilities of each case and findings from various modal analyses. Finally, concluding remarks are given in Section \ref{sec:concl}.

\section{Methods and Setup} 
\label{sec:methodSetup}

\subsection{Physical Model Problem}
The supersonic nozzle flow at the exit of the engine of figure \ref{fig:engine} is modeled as shown in figure~\ref{fig:nozzle}, and includes the splitter plate with the aft-deck below and the SERN surface above. The nozzle dimensions are summarized in table~\ref{tab:nozz_dimen}. All length scales in the subsequent discussion are non-dimensionalized by the splitter plate width $W$, where $W$ corresponds to an experimental length scale of 82.296 mm.
Further details on the geometry information may be found in the works by \cite{simmons2009design}, \cite{berry2016investigating}, \cite{magstadt2017investigating}, and \cite{stack2019turbulence}. 
\begin{table}
    \begin{center}
    \def~{\hphantom{0}}
        \begin{tabular}{P{5cm} P{3cm} P{3cm}}
            Nozzle geometry parameter     & Description              & Value   \\ [1pt]
            \midrule
            $\delta_{\text{SP}}/W$        & Splitter plate thickness & 0.039   \\ [1pt]
            $h_{\text{t}}/W$              & Throat height            & 0.229   \\ [1pt]
            $h_{\text{NE}}/W$             & Nozzle exit height       & 0.357   \\ [1pt]
            $L_{\text{SP}}/W$             & Splitter plate length    & 0.190   \\ [1pt]
            $L_{\text{ER}}/W$             & Expansion ramp length    & 0.830   \\ [1pt]
            $L_{\text{AD}}/W$             & Aft-deck length          & 1.094   \\ \hline
        \end{tabular}
    \caption{Nozzle dimensions (according to figure \ref{fig:nozzle}(b)) normalized by the splitter plate width $W$. }
    \label{tab:nozz_dimen}
    \end{center}
\end{table}

As noted earlier, among the three streams produced by the engine, the core and fan streams (figure~\ref{fig:engine}) are well mixed before entering the SERN, forming one uniform main stream entering the nozzle (figure~\ref{fig:nozzle}(a)). In the numerical setup, only two physical streams are simulated for the nozzle flow: a supersonic main stream and a sonic bypass stream, which are separated by a splitter plate. The two streams mix before passing through and exiting the nozzle. The operating conditions of each stream are the same as those of prior experimental and numerical studies by \cite{berry2016investigating}, \cite{magstadt2017investigating}, and \cite{stack2019turbulence}. The engine core flow operates at a nozzle pressure ratio (NPR= $P_\text{total}/P_\text{ref}$) of NPR$_\text{core}$ = 4.25, while the bypass stream operates at NPR$_\text{bypass}$ = 1.89, corresponding to Mach numbers of 1.6 and 1.0, respectively, based on the isentropic relation. The nozzle temperature ratio (NTR= $T_\text{total}/T_\text{ref}$) is set to unity for both flows, and is representative of the unheated jet utilized in the experiment by \cite{berry2016investigating} and \cite{magstadt2017investigating}.

\subsection{Active Flow Control Configuration}
\begin{figure}
    \centering
    \includegraphics[width=0.65\textwidth]{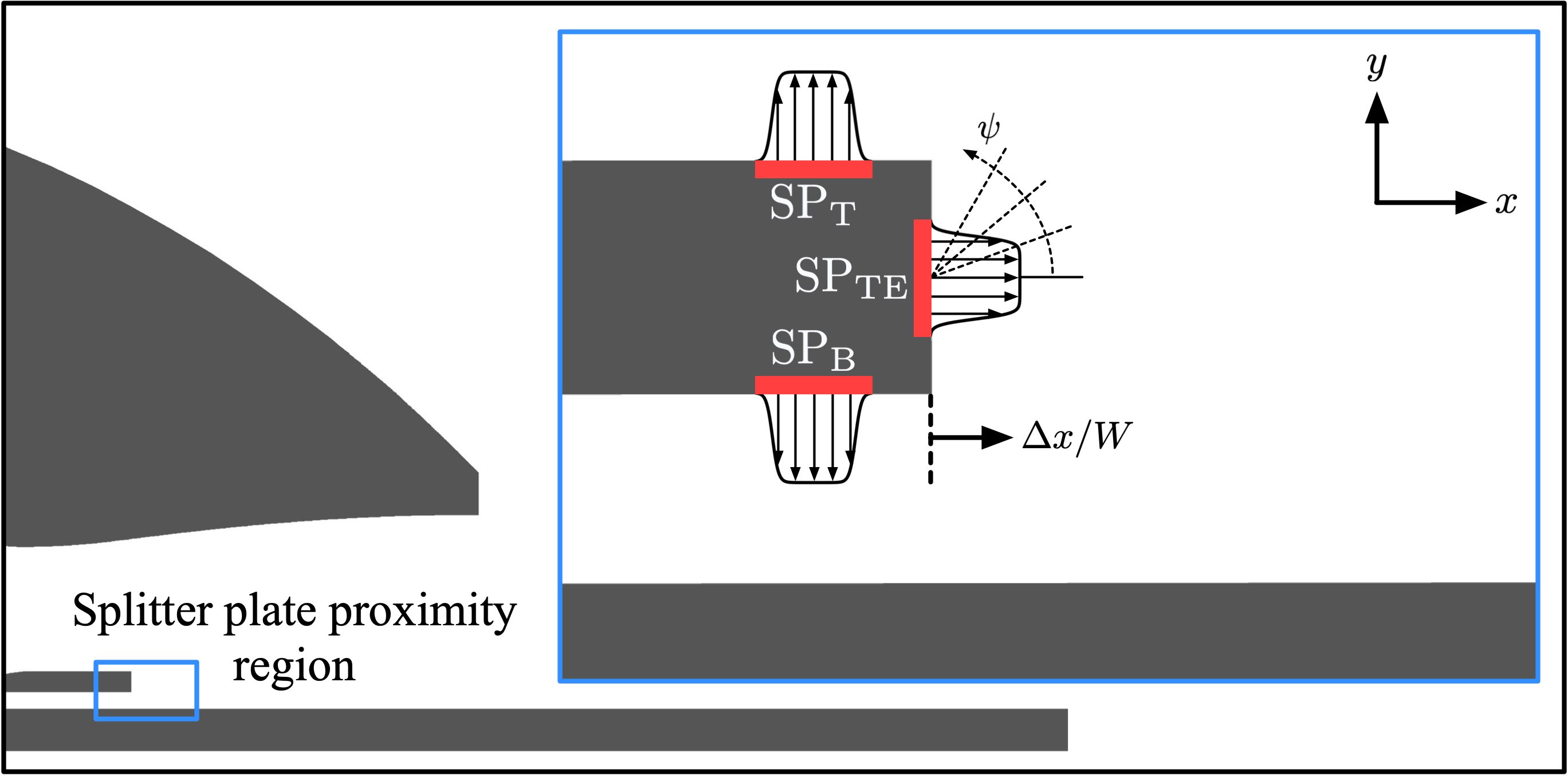}
    \caption{Schematic of the active flow control configuration. Each slotted actuation surface is indicated with a thick red line, \includegraphics[width=0.05\textwidth]{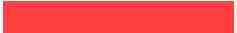}. $\text{SP}_\text{T}$ = splitter plate top surface, $\text{SP}_\text{TE}$ = splitter plate trailing edge surface, $\text{SP}_\text{B}$ = splitter plate bottom surface.}
    \label{fig:afc_config}
\end{figure}

Active flow control in the form of micro-jet actuation is introduced at different locations on the splitter plate near its trailing edge. Jets in crossflow (JICF) have been extensively studied (\cite{li2020effect, liu2018structures, karagozian2014jet, cortelezzi2001formation}) and used in active control efforts in a multitude of supersonic fluid flow problems (\cite{liu2019flow, guo2018parametric, guo2017numerical, karagozian2010transverse}). The specific control objectives of the present work are to suppress the resonant tone, attenuate shock strength, and reduce the surface loading on the aft-deck caused by the intense flow unsteadiness. Figure~\ref{fig:afc_config} depicts the details of the control configuration. Three locations are considered independently, in which the actuator is placed on either the splitter plate top surface ($\text{SP}_\text{T}$), or the splitter plate bottom surface ($\text{SP}_\text{B}$), or the vertical surface of the splitter plate trailing edge ($\text{SP}_\text{TE}$). These choices are motivated by the resolvent analysis, discussed below, which indicated that this region is most receptive to external disturbances.

The micro-jet actuator is characterized by two parameters: the location and angle of injection, $\psi$. At each location, steady-blowing is introduced at different injection angles over a range of $\psi\in[-60\degree, 90\degree]$, where $\psi$ is measured counter-clockwise from the stream-wise direction, $x$. The momentum coefficient, $C_\mu$, and the jet-to-mainstream momentum flux ratio, $J$, are defined as
\begin{equation}
    C_\mu = \frac{\dot m_\text{a} V_\text{a}}{\frac{1}{2}\overline{\rho}_\text{e} \overline{U}_{\text{e}}^2 A_\text{e}},\text{~~~and~~~}
    J = \frac{\rho_\text{a}V_\text{a}^2}{\overline{\rho}_\text{e}\overline{U}_{\text{e}}^2},
\end{equation}
where $\dot m_\text{a}$ is the actuation's blowing mass flow rate, $V_\text{a}$ is the center local jet velocity, the sign of $\overline{(\cdot)}$ denotes spatial- and time-averaged quantity, $\rho_\text{e}$ is the density at the nozzle exit, $U_\text{e}$ is the velocity at the nozzle exit, $A_\text{e}$ is the area of the nozzle exit, and $\rho_\text{a}$ is the density of the micro-jet. 
A hyperbolic-tangent velocity profile is prescribed for all the control cases, as shown in figure \ref{fig:afc_config}, with a center local Mach number of 1.0 at the slot. 
The actuator location, injection angle, momentum coefficient, and jet-to-mainstream momentum flux ratio of all the control configurations considered in the present work are tabulated in table~\ref{tab:afc_cases}. 
\begin{table}
    \begin{center}
        \begin{tabular}{P{2cm} P{2cm} P{5cm} P{1.5cm} P{2cm}}
            Location                & $\Delta x/W$    & $\psi$                                                      & $C_\mu$                & ~~~~ $J$ \\ \hline
            $\text{SP}_\text{T}$    & $-0.019$        & $90$\degree, $60$\degree, $45$\degree, $30$\degree            & 0.090                  & ~~~~ 0.831 \\           
            $\text{SP}_\text{TE}$   & $-$             & $-60$\degree, $-45$\degree, $-30$\degree, $0$\degree, $30$\degree, $45$\degree, $60$\degree    & 0.090                  & ~~~~ 0.831 \\     
            $\text{SP}_\text{B}$    & $-$0.019        & $-30$\degree, $-45$\degree, $-60$\degree                      & 0.090                  & ~~~~ 0.831 \\           
        \end{tabular}
    \caption{Summary of the active control configurations.$\text{SP}_\text{T}$, $\text{SP}_\text{TE}$, and $\text{SP}_\text{B}$  = splitter plate top, trailing edge, and bottom surfaces. $\psi$: injection angle, $\Delta x/W$: stream-wise distance from actuation center to splitter plate trailing edge, $C_\mu$: momentum coefficient, and $J$: jet-to-mainstream momentum flux ratio.}
    \label{tab:afc_cases}
    \end{center}
\end{table}

\subsection{Numerical Model}
\begin{figure}
    \centering
    \includegraphics[width=0.7\textwidth]{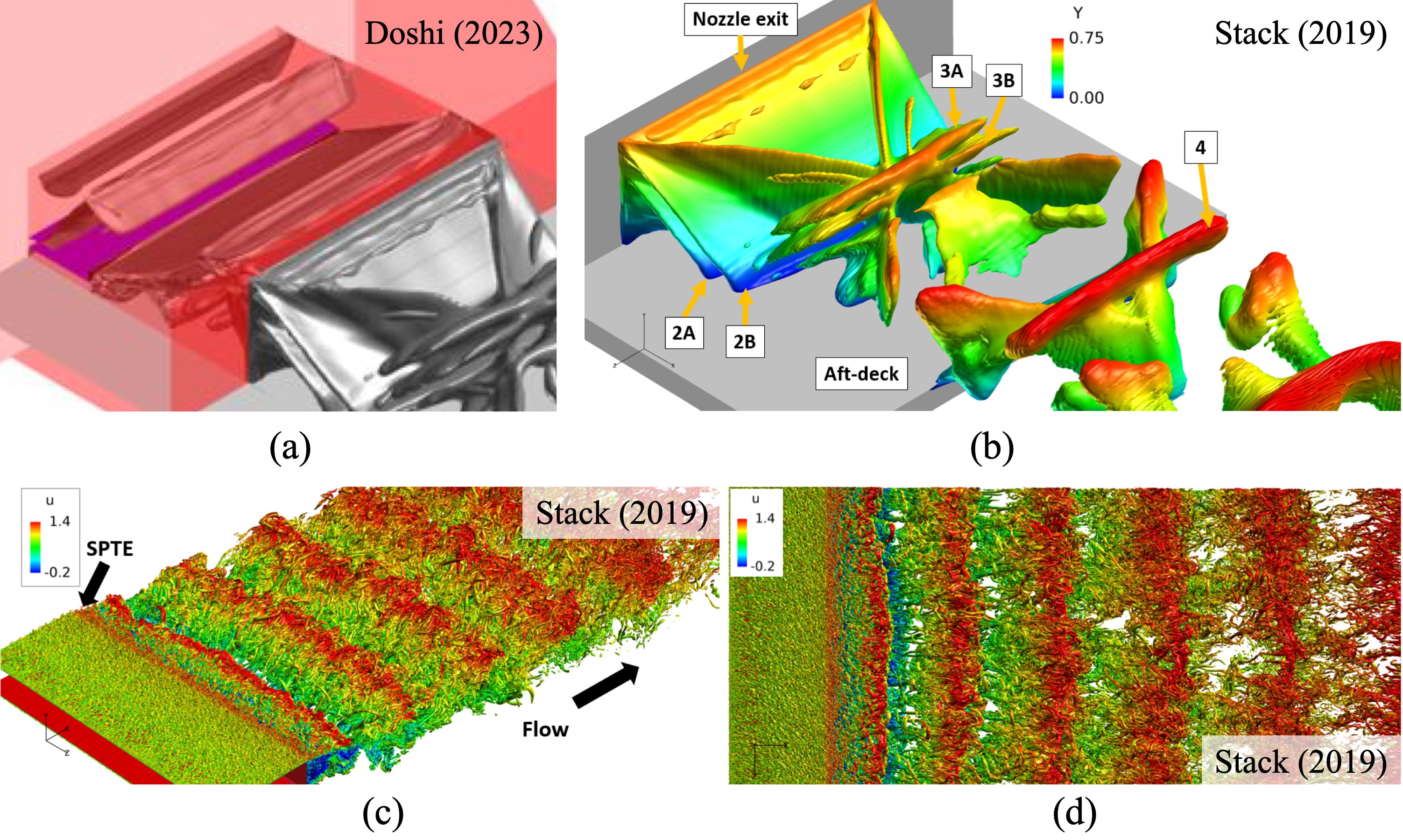}
    \caption{3-D simulation results by \cite{doshi2023modal} and \cite{stack2019turbulence}. Iso-surfaces of (a) $\bm{U} \cdot \nabla P = 0.3$ and (b) $\bm{U} \cdot \nabla P = 0.2$ for the full SERN configuration. (c-d) Instantaneous visualization of the Q-criterion for an isolated shear layer.}
    \label{fig:2Dstructs}
\end{figure}

In prior numerical studies, the dominant flow physics within the nozzle, except very near the sidewalls, were found to be largely two-dimensional (2-D). Figure \ref{fig:2Dstructs} depicts the three-dimensional (3-D) simulations performed by \cite{doshi2023modal} and \cite{stack2019turbulence}, where frames (a-b) illustrate the shock topology of the full SERN configuration. Within the nozzle, the shock structures are observed to be largely two-dimensional, while shocks originating from the corner regions at the nozzle exit result in the formation of a three-dimensional shock train. Since the high-frequency tone associated with this jet was found to originate from the mixing of the two incoming flows within the nozzle, a 3-D simulation of the isolated shear layer was performed and is shown in figure \ref{fig:2Dstructs}(c-d). It was found that the coherent structures associated with the shedding instability were also primarily two-dimensional. Since we seek to perform a relatively large number of parametric studies, a 2-D rectangular domain is used to simulate the flow.

Simulations are performed using the solver, \textit{CharLES} (\cite{bres2019modelling, bres2018importance}). The solver uses a second-order finite-volume method and a third-order Runge-Kutta temporal scheme to solve the compressible Navier--Stokes equations, and the relative-solution-ENO scheme is utilized to capture shocks. The reference values used in the simulations are $\rho_{\text{ref}}$ = 1.0, $P_{\text{ref}}$ = 0.714, and $T_{\text{ref}}$ = 1.0, which yields the reference speed of sound $c_\text{ref} = 1$. 
This corresponds to the dimensional reference values used by \cite{stack2019turbulence}, which are $\rho_{\text{ref}} = 1.173~kg/m^3$, $c_{\text{ref}} = 347.189~m/s$, $P_{\text{ref}} = \rho_{\text{ref}} c_{\text{ref}}^2$, and $T_{\text{ref}} = 300~K$.
The Reynolds number is of $Re = \rho_\infty U_\text{jet} D_h/\mu_\infty = 1.5\times10^5$, where $\rho_\infty$ is the freestream density, $U_\text{jet} = 1.6$, calculated based on the isentropic relation and NPR$_\text{core} = 4.25$, $D_h$ is the hydraulic diameter, and $\mu_\infty$ is the freestream dynamic viscosity. The hydraulic diameter, $D_h$, is calculated based on the rectangular nozzle exit, defined as
\begin{equation}
    D_h = \frac{2h_{\text{NE}}W}{h_{\text{NE}}+W},
\end{equation}
where $h_\text{NE}$ and $W$ are the nozzle exit height and width, respectively. Here, the nozzle width is equivalent to the splitter plate width.

\begin{figure}
    \centering
    \includegraphics[width=1\textwidth]{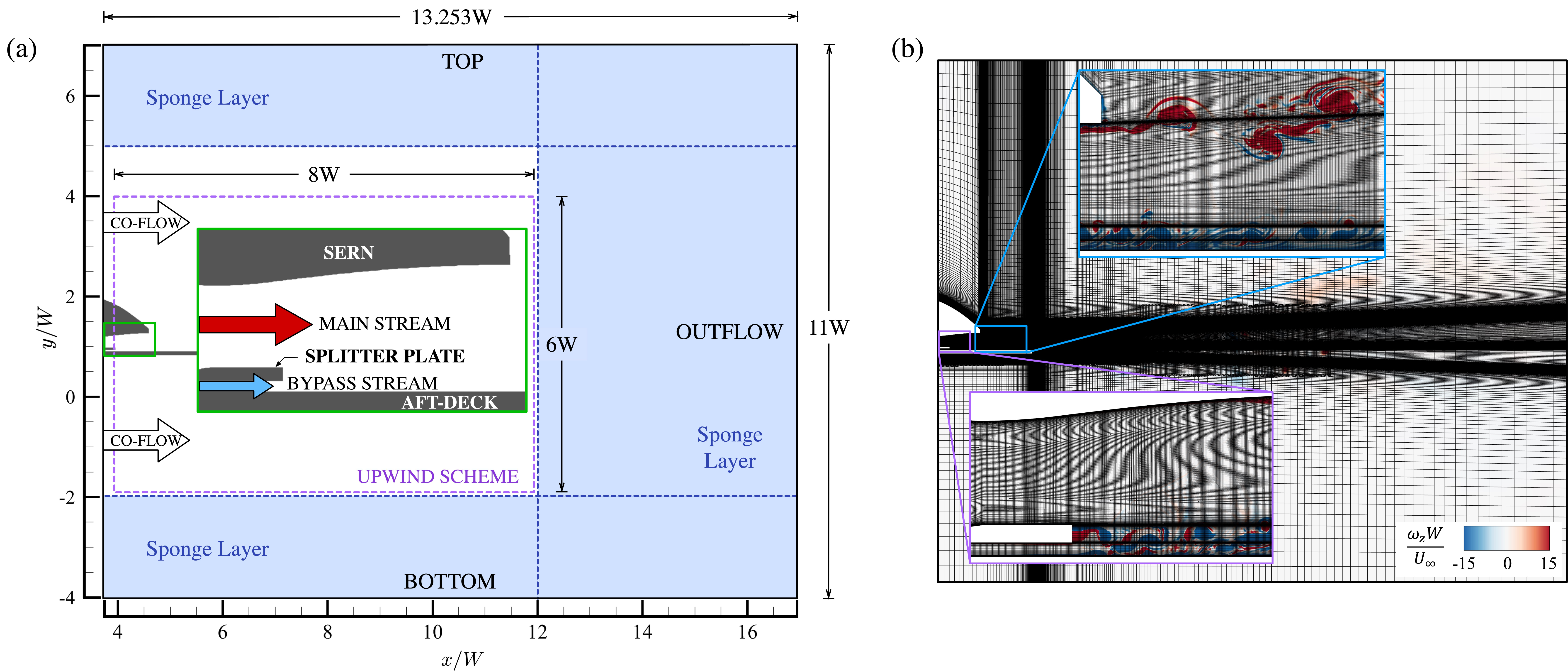}
    \caption{(a) Schematic of computational domain setup and flow configuration. Nozzle geometry is shown in gray, and sponge layers are indicated in blue. (b) Grid topology with instantaneous vorticity.}
    \label{fig:config_mesh}
\end{figure}

\begin{figure}
    \centering
    \includegraphics[width=0.95\textwidth]{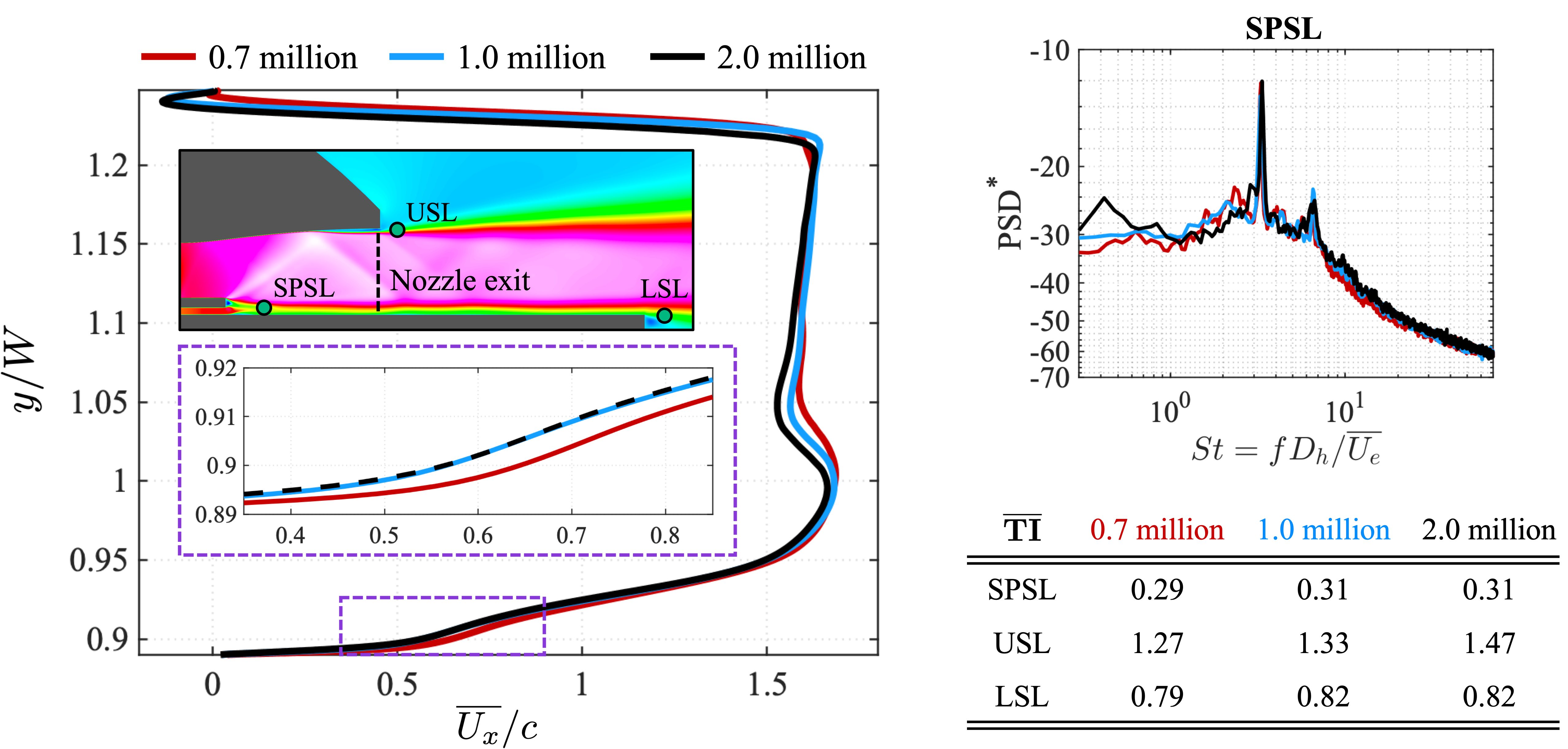}
    \caption{Time-averaged velocity profiles at the nozzle exit, pressure spectra, and turbulence intensities for three grid resolutions: 0.7, 1.0, and 2.0 million. The case with 2.0 million cells is shown dashed for clarity in the smaller window.}
    \label{fig:GRS}
\end{figure}

Figure \ref{fig:config_mesh}(a) shows the computational domain and flow configuration used in the present study.
The length and width of the domain are 13.253$W$ and 11$W$, respectively. The stream-wise and wall-normal directions are denoted by $x$ and $y$ with corresponding velocity components of $u$ and $v$, where a Cartesian coordinate system is used. As shown in figure~\ref{fig:config_mesh}(b), a structured mesh with non-uniform spacing is used, where grids along no-slip walls and around primary jet flows are refined to resolve boundary layers and small-scale structures. 
A grid resolution study is performed for three mesh sizes containing 0.7 million, 1.0 million, and 2.0 million points, where each grid further refines the region surrounding the primary jet flow. The time-averaged velocity profiles at the nozzle exit, pressure spectra, and turbulence intensities (TI) of the three shear layers are compared between each grid in figure~\ref{fig:GRS}. The TI is calculated using the velocity components as
\begin{equation}
    \text{TI}(t) = \sqrt{\frac{\frac{1}{2}(u'(t)^2 + v'(t)^2)} {\overline{u}^2 + \overline{v}^2}},
\end{equation}
where $u'(t)$ and $v'(t)$ are the time-varying fluctuating components of $u$ and $v$, respectively. While all grids are able to capture the resonant tone, the difference in TI between the 1.0 million and 2.0 million grids is small. Moreover, the corresponding velocity profiles agree well near the walls. The 1.0 million mesh is chosen in favor of the computational cost and is used for the remainder of this study (figure~\ref{fig:config_mesh}(b)).

A co-flow of Mach number $M$ = 0.01 is employed at the left boundary to mimic the experimental setup used by \cite{berry2016investigating} and \cite{magstadt2017investigating} at the Skytop Turbulence Lab at Syracuse University. Characteristic boundary conditions are applied at the main and bypass stream boundaries. The inflow conditions are determined using the designed nozzle pressure ratio and nozzle temperature ratio for both streams, and are prescribed to be uniform flows. All the walls of the SERN, splitter plate, and aft-deck are assigned to be adiabatic with a no-slip condition. An upwind scheme (\cite{bres2017unstructured}) is employed in a region of length 8$W$ and width 6$W$. Sponge layers are placed outside of the upwind domain along the top, bottom, and outflow boundaries. In this region, a source term is used to damp outgoing waves and prevent reflection from boundaries into the domain (figure \ref{fig:config_mesh}(a)).

\begin{figure}
    \centering
    \includegraphics[width=0.9\textwidth]{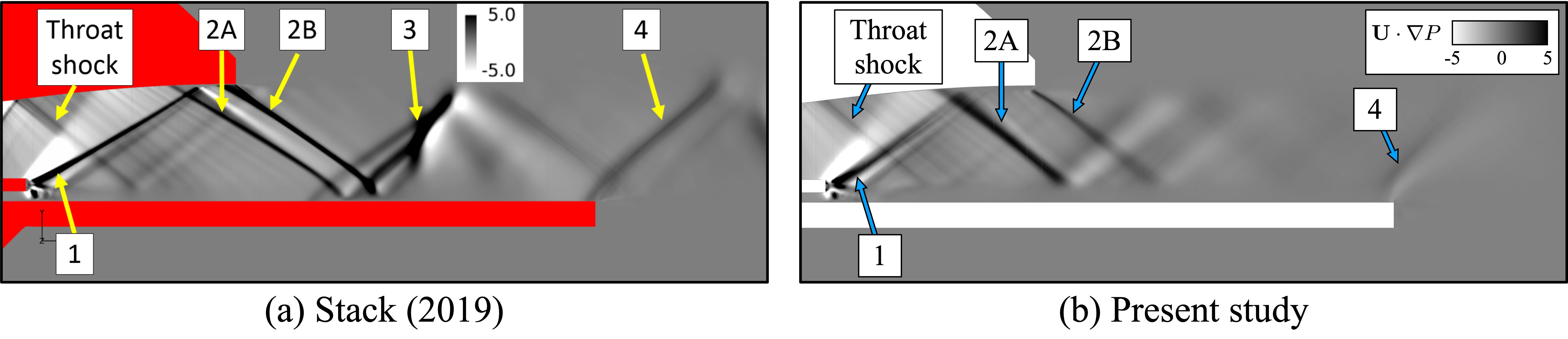}
    \caption{Comparison between the time-averaged $\bm{U}\cdot \nabla P$ from the 3-D simulation by (a) \cite{stack2019turbulence} and (b) 2-D results in the present work.}
    \label{fig:2Dvs3D}
\end{figure}

A comparison between the 3-D simulation performed by \cite{stack2019turbulence} and the current 2-D results is shown in figure \ref{fig:2Dvs3D}. Similar shock train components as captured by \cite{stack2019turbulence} are observed in the 2-D simulation, most notably Shocks~1, 2A, 2B, and 4. The effects of choices made to enable parametric studies have a modest effect on the solution.
Thus,  Shock~1 in the 3-D simulation impinges along the SERN closer to the nozzle lip since Stack's Reynolds number is one order of magnitude higher ($Re_{D_h} = 2.70\times 10^6$) than the current 2-D study.
In turn, Shock~2A develops further upstream in the present work, and the reflected shocks from 2A and 2B do not coalesce into Shock~3, which also consists of the 3-D corner shocks, as shown in figure \ref{fig:2Dstructs}(b).
However, Shocks~2B and 4 are observed in the same locations for both the 2-D and 3-D configurations.
The lowering of Reynolds number and use of two-dimensional simulations  substantially reduces the required computational resources without compromising the  primary mechanisms of interest behind the development of Shocks~1 and 2A. 
In fact, the characteristic high-frequency tone captured in the present work is the same as that reported in simulation and experiment, as will be discussed later.

\subsection{Spectral Proper Orthogonal Decomposition}
\begin{figure}
 \centering
     \includegraphics[width=0.5\textwidth]{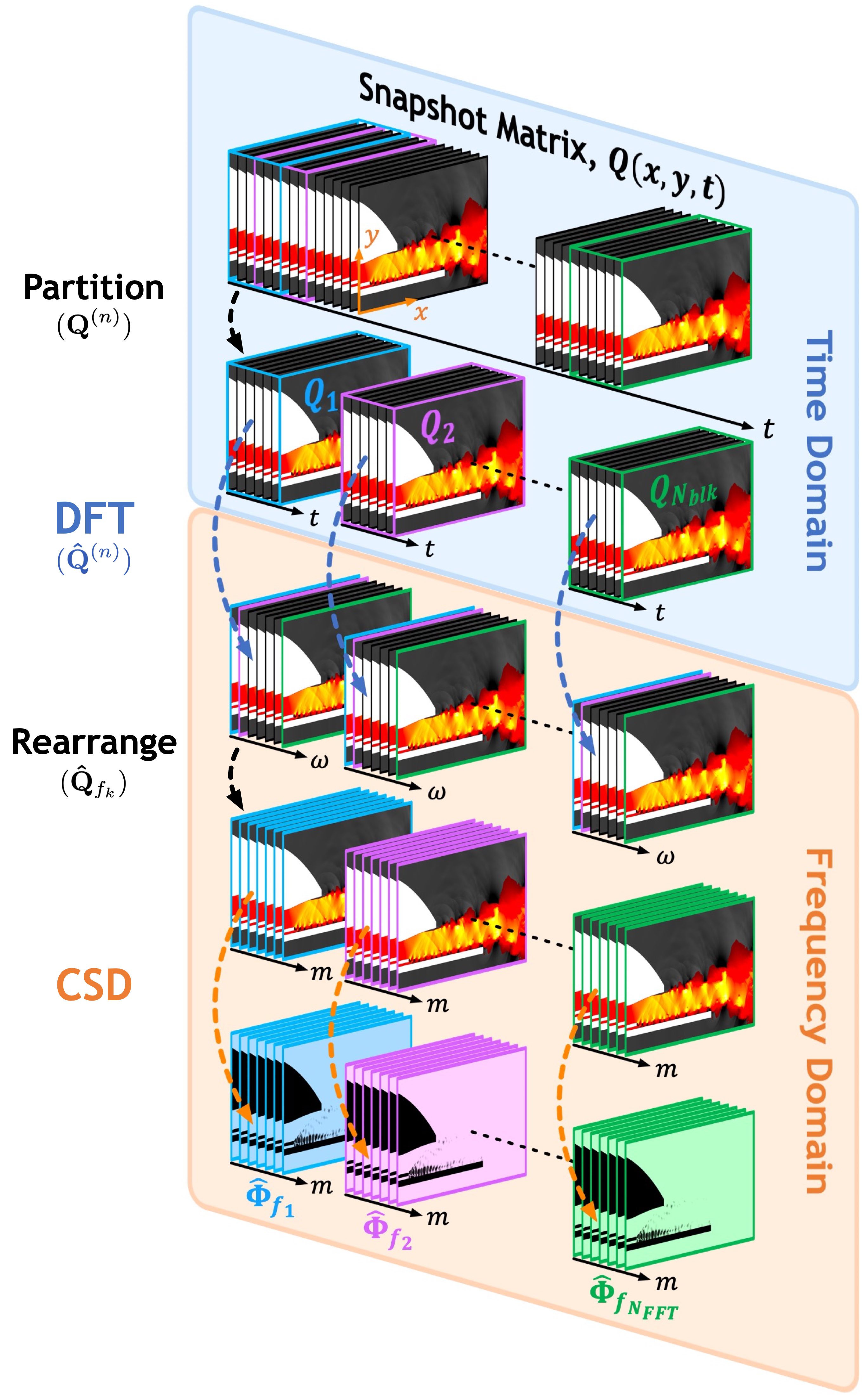}
     \caption{Schematic of SPOD algorithm. Each slice in $\bm{Q}$ represents a snapshot in time.}
     \label{fig:spod-al}
 \end{figure}

To gain insight into the complex flow dynamics, the spectral proper orthogonal decomposition algorithm, a method initiated by \cite{glauser1987coherent} and analyzed recently by \cite{towne2018spectral}, is applied on the baseline and representative control cases to extract coherent flow structures associated with prominent frequencies. In the SPOD algorithm shown in figure~\ref{fig:spod-al}, an instantaneous state of the flow $\bm{q}(\bm{x},t)$ at time $t_k$ can be rearranged into a column vector $\bm{q}_k \in \mathbb{R}^N$, where $N$ is equal to the number of data points times the number of flow variables. Given $M$ equally spaced snapshots, a large data matrix $\bm{Q}$ is constructed such that $\bm{Q} = [\bm{q}_1 ~~ \bm{q}_2 ~ \cdots ~ \bm{q}_M] \in \mathbb{R}^{N \times M}$. This snapshot matrix is then partitioned into overlapping blocks $\bm{Q}^{(n)} = [\bm{q}_1^{(n)} ~ \bm{q}_2^{(n)} ~ \cdots ~ \bm{q}_{N_{\text{FFT}}}^{(n)}] \in\mathbb{R}^{N \times N_{\text{FFT}}}$ with each block containing $N_{\text{FFT}}$ snapshots. A Discrete Fourier Transform is then applied on each block to obtain $\hat{\bm{Q}}^{(n)} = [\hat{\bm{q}}_1^{(n)} ~ \hat{\bm{q}}_2^{(n)} ~ \cdots ~ \hat{\bm{q}}_{N_{\text{FFT}}}^{(n)}]$ with $\hat{\bm{q}}_k^{(n)}$ being the Fourier component at frequency $f_k$ in the $n$th block. The Fourier coefficients at each frequency $f_k$ from each block are rearranged into a new data matrix $\hat{\bm{Q}}_{f_k} = \sqrt{\kappa}[\hat{\bm{q}}_k^{(1)} ~ \hat{\bm{q}}_k^{(2)} ~ \cdots ~ \hat{\bm{q}}_{k}^{(N_{\text{blk}})}]$ where $\kappa = \Delta t/(sN_{\text{blk}})$. Here, $N_{\text{blk}}$ is the number of blocks and $s = \sum\nolimits_{j=1}^{N_{\text{FFT}}} w_j^2$, where the scalar weights $w_j$ can be used to reduce spectral leakage due to the non-periodicity in each block. The cross-spectral density tensor at frequency $f_k$ can then be estimated as $\bm{S}_{f_k} = \hat{\bm{Q}}_{f_k} \hat{\bm{Q}}_{f_k}^*$. An eigenvalue decomposition is then performed on an $N \times N$ matrix, 
\begin{equation}
    \bm{S}_{f_k} \bm{\Gamma \Phi}_{f_k} = \bm{\Phi}_{f_k} \bm{\Lambda}_{f_k},
\end{equation}
with $\bm{\Phi}_{f_k}$ being the SPOD modes that are ranked by the corresponding eigenvalues given by the diagonal matrix $\bm{\Lambda}_{f_k}$. It is noted that the choice of norm is expressed through the discrete weight matrix $\bm{\Gamma}$. A detailed guide to the SPOD algorithm can be found in the works by \cite{towne2018spectral} and \cite{schmidt2020guide}.

In the present work, SPOD is computed using the density variable, where the simulation results are interpolated onto a $500\times 250$ uniform grid prior to performing SPOD. A grid study has been performed to ensure the SPOD results were not grid dependent. The weight matrix $\bm{\Gamma}$ is assigned to be an identity, which directly measures the variance of the data and is the natural choice of the norm if all data points are considered to be equally significant (\cite{schmidt2020guide}). For all the SPOD analysis performed in the present work, 5{,}000 consecutive snapshots covering a time period of $t \overline{U_e}/D = 2.07$ are used, where $D$ is the stream-wise length of the computational domain shown in figure \ref{fig:config_mesh}. Each snapshot is separated by $\Delta t \overline{U_e}/D = 4.14 \times 10^{-4}$. The snapshot matrix $\bm{Q}$ is partitioned into $N_{\text{blk}} = 15$ blocks with 75$\%$ overlap such that each block contains $N_{\text{FFT}} = 1{,}111$ snapshots. 

\subsection{Resolvent Formulation}
To gain physical insights into active control configurations, resolvent analysis is performed using the turbulent mean flow to find optimal energy amplifications. Comprehensive details of the resolvent formulation can be found in the works by \cite{mckeon2010critical}, \cite{sun2020resolvent}, and \cite{yeh2019resolvent}. Here, the compressible Navier-Stokes equations are expressed as
\begin{equation}
    \frac{\partial \bm{q}}{\partial t} = \mathcal{N}(\bm{q}),
    \label{eqn:compNS}
\end{equation}
where $\mathcal{N}$ is the nonlinear Navier-Stokes operator applied on the state variables $\bm{q} = [\rho,u,v,w,T]^T$.  Here $\rho$ is the density, $u, v, w$ are the velocity components, and $T$ is the temperature. The state variables can be Reynolds decomposed into $\bm{q}(x,y,z,t) = \overline{\bm{q}}(x,y) ~ + ~\bm{q}'(x,y,z,t)$, where $\overline{\bm{q}}$ and $\bm{q}'$ are the mean and fluctuating components, respectively. Substituting the decomposed state variables into Equation (\ref{eqn:compNS}) yields
\begin{equation}
    \frac{\partial \bm{q}'}{\partial t} = \mathcal{L}_{\overline{\bm{q}}}(\bm{q}') + \bm{f}',
    \label{eqn:lnzdNS}
\end{equation}
where $\mathcal{L}_{\overline{\bm{q}}}$ is the linearized Navier-Stokes operator about a base state and $\bm{f}$ denotes higher-order nonlinear terms or external forcing. In the present work, the operator $\mathcal{L}_{\overline{\bm{q}}}$ is constructed using a time-averaged flow. The terms $\bm{q}'$ and $\bm{f}'$ are then represented as a sum of Fourier modes with 
\begin{equation}
        \bm{q}'(x,y,z,t) = \hat{\bm{q}}_{\omega,\beta}(x,y)e^{i(\beta z-\omega t)}
\end{equation}
and
\begin{equation}
        \bm{f}'(x,y,z,t) = \hat{\bm{f}}_{\omega,\beta}(x,y)e^{i(\beta z-\omega t)},
\end{equation}
where $\omega$ is a complex-valued frequency and $\beta$ is a real-valued spanwise wavenumber. Substituting the Fourier expansions into Equation~(\ref{eqn:lnzdNS}) brings the linearized Navier-Stokes equations to Fourier space as
\begin{equation}
    -i\omega~\hat{\bm{q}}_{\omega,\beta} = \mathcal{L}_{\overline{\bm{q}}}(\hat{\bm{q}}_{\omega,\beta} ;\beta) + \hat{\bm{f}}_{\omega,\beta},
    \label{eqn:fsNS}
\end{equation}
where Equation~(\ref{eqn:fsNS}) can be rearranged and expressed as
\begin{equation}
    \hat{\bm{q}}_{\omega,\beta} = [-i\omega \bm{I} - \mathcal{L}_{\overline{\bm{q}}}(\beta)]^{-1} ~ \hat{\bm{f}}_{\omega,\beta}.
\end{equation}
Here, 
\begin{equation}
    \mathcal{H}_{\overline{\bm{q}}}(\omega,\beta) = [-i\omega \bm{I} - \mathcal{L}_{\overline{\bm{q}}}(\beta)]^{-1}
    \label{eqn:rslvtOp}
\end{equation}
is referred to as the resolvent operator and acts as a transfer function between the input $\hat{\bm{f}}$ and output $\hat{\bm{q}}$. Performing a singular value decomposition (SVD) on the operator $\mathcal{H}_{\overline{\bm{q}}}(\omega,\beta)$ yields optimal forcing-response mode pairs with an associated gain, evaluated as
\begin{equation}
    \mathcal{H}_{\overline{\bm{q}}}(\omega,\beta) = \bm{\mathcal{Q}} \Sigma \bm{\mathcal{F}}^*
\end{equation}
where the singular vectors $\bm{\mathcal{Q}}$ and $\bm{\mathcal{F}}$ represent the response and forcing modes, respectively. The corresponding gains for each forcing-response pair are ranked in terms of their magnitude as $\Sigma = \text{diag}(\sigma_1,~\sigma_2,~\cdots~,\sigma_m$). In the present study, the spanwise wavenumber is always chosen to be $\beta = 0$ such that focus is placed on the two-dimensional mechanisms, and Chu's norm for compressible flows (\cite{chu1965energy}) is used for the resolvent analysis (\cite{yeh2019resolvent, towne2018spectral}). 

In the event of an unstable base state, the stability of the linearized Navier-Stokes operator $\mathcal{L}_{\overline{\textbf{q}}}$ is analyzed as a precursor to the resolvent analysis (\cite{sun2017biglobal, sun2020resolvent, liu2021unsteady}). An eigenvalue decomposition of the linear operator is performed as
\begin{equation}
    \omega \hat{\bm{q}}_{\omega,\beta} = i\mathcal{L}_{\overline{\textbf{q}}}(\beta) \hat{\bm{q}}_{\omega,\beta},
\end{equation}
where $\omega$ is a complex-valued eigenvalue with the real $\omega_r$ and the imaginary $\omega_i$ components representing the frequency and growth rate, respectively. The most unstable eigenvalue of the system is utilized to determine a real-valued parameter $\alpha$, where $\alpha$ is chosen such that $\alpha > \text{max}(\omega_i)$. This parameter introduces an exponential discount to the system, forming the discounted operator presented by \cite{jovanovic2004modeling}. This limits the flow response to a finite time window characterized by $1/\alpha$, and focus is placed on the short-term dynamics. Introducing the discount parameter $\alpha$ into the original resolvent operator presented in Equation (\ref{eqn:rslvtOp}) yields 
\begin{equation}
    \begin{split}
        \mathcal{H}_{\overline{\bm{q}},\alpha}(\omega,\beta) & = [-i\omega \bm{I} - \{\mathcal{L}_{\overline{\textbf{q}}}(\beta) - \alpha\bm{I}\}]^{-1} \\
        & = [-i(\omega+i\alpha)\bm{I} - \mathcal{L}_{\overline{\textbf{q}}}(\beta)]^{-1}
    \end{split}
\end{equation}
and $\mathcal{H}_{\overline{\bm{q}},\alpha}$ is referred to as the discounted resolvent operator. It is noted that choosing $\alpha = 0$ recovers the original formulation shown in Equation (\ref{eqn:rslvtOp}), which represents a scenario in which an infinite-long time window is considered.

\section{Results}
\label{sec:results}

\subsection{Baseline Flow Features} 
In this section, the instantaneous and mean flow features of the baseline case are discussed. Spectral analysis is performed to identify and extract coherent structures that correspond to prominent frequencies. The dominant instabilities of the flow are investigated, and the discounted resolvent analysis is applied to gain insight into the forcing-response relations of the flow, such that optimal locations for introducing perturbations can be discovered.

\subsubsection{Instantaneous and mean flow features}
\begin{figure}
\centering
    \includegraphics[width=1\textwidth]{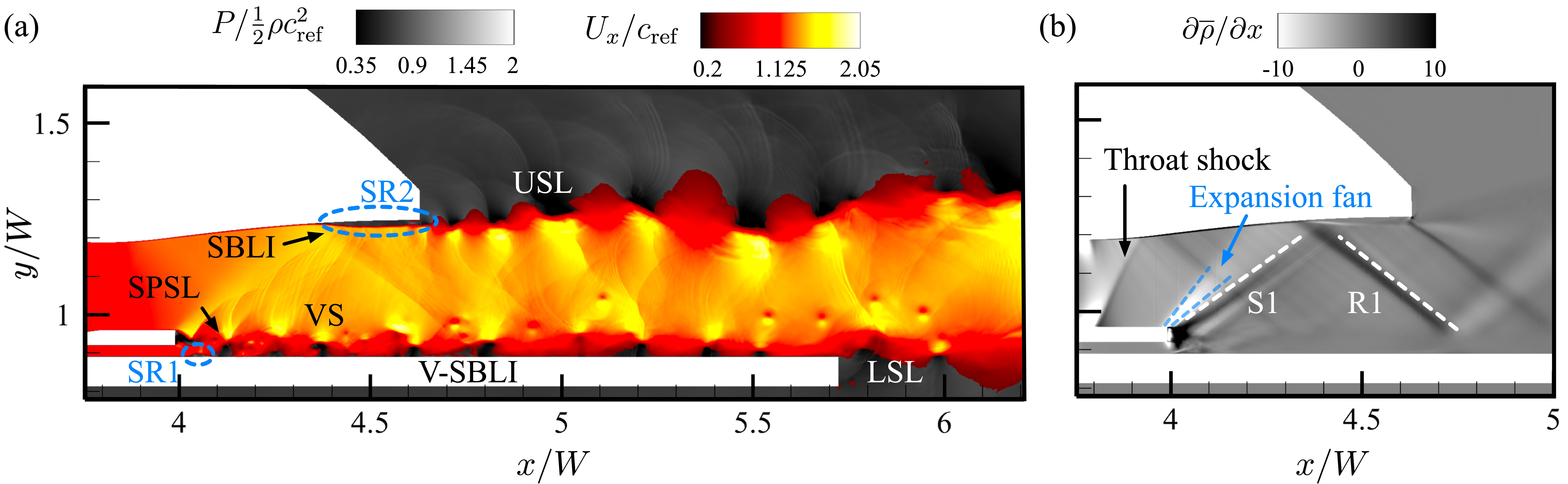}
    \caption{Instantaneous (a) and mean (b) flow fields for baseline case.}
    \label{fig:base_flow}
\end{figure}

An instantaneous and mean flow field of the baseline case is shown in figure~\ref{fig:base_flow}.  Frame~(a) displays shaded color contours of the stream-wise velocity that are visualized together with black-white contours of the pressure field, while figure~\ref{fig:base_flow}(b) shows the time-averaged density gradient in the $x$-direction so that negative and positive values highlight expansions and shocks, respectively, in the supersonic flow. As shown in figure~\ref{fig:base_flow}(b), an expansion fan forms as the main stream turns clock-wise around the SPTE. Following this, a strong primary shock (S1) is observed, whose structure is comprised of an oblique shock and vortex-induced compression waves. The former is formed as the main stream aligns with the horizontal direction downstream of the SPTE, while the latter is associated with vortex shedding (VS) generated due to the mixing between the main and bypass streams. These affect the properties of the initial turning-associated oblique shock to form the S1 shock. As shown in figure~\ref{fig:base_flow}(a), mixing of the main-bypass streams additionally causes a small separation region after the SPTE (SR1). The shedding vortices generated by the mixing of the two streams impinge on the aft-deck and convect downstream, forming a splitter plate shear layer (SPSL). The S1 shock travels across the core flow and impinges on the expansion ramp of the nozzle, inducing a shock-boundary-layer interaction (SBLI) and causing a shock-induced flow separation (SR2). The reflected R1 shock then impinges on the aft-deck, where it encounters the incoming vortices, and a vortex-shock-boundary-layer interaction occurs (V-SBLI). A shock is observed emanating from the nozzle lip due to the static pressure at the nozzle exit being lower than the ambient, indicative of the overexpanded nature of the baseline configuration. This is consistent with prior experimental and numerical studies (\cite{berry2016investigating, magstadt2017investigating, stack2019turbulence}).
When the flow exits the nozzle, the upper and lower shear layers (USL and LSL) form with the ambient. 

\subsubsection{Spectral analysis}
\begin{figure}
    \centering
    \includegraphics[width=0.55\textwidth]{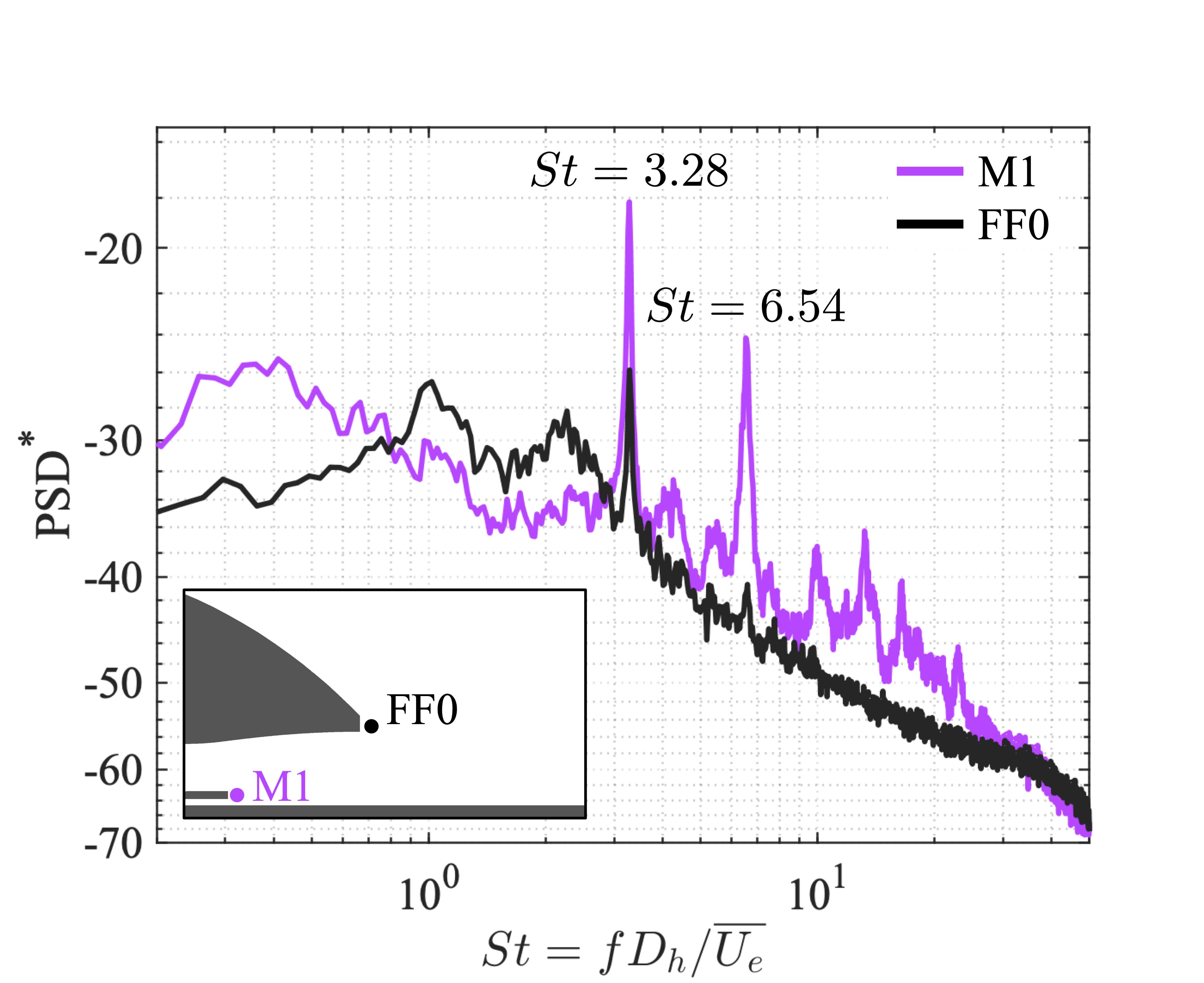}
    \caption{Power spectral density of the pressure signal from two point probes in the baseline flow. Probe locations are shown in the lower left-hand corner.}
    \label{fig:base_psd}
\end{figure}

The frequency spectrum is first examined using the power spectral density (PSD) of the pressure time series at various point probes. The PSD is calculated using Welch's method (\cite{welch1967use}) with a Hanning window and 75\% overlap. The data is sampled covering the entire data length of $t_D = t\overline{U_{\text{e}}}/D > 10$ for all data sets, and the sampling frequency is $f_s = D/\Delta t\overline{U_\text{e}} = 1.0526\times10^4$. Figure~\ref{fig:base_psd} presents a non-dimensional power spectral density $\text{PSD}^* = 10\text{log}_{10} (P_{xx} U_\text{jet}/q_\infty^2 D_h)$ for two point probes, denoted as M1 and FF0, as well as their corresponding locations in the flow field, where $P_{xx}$ is the autospectral density and $q_\infty$ is the freestream dynamic pressure. The M1 probe $(x,y)/W = (4.017, 0.940)$ is placed in the splitter plate wake and directly captures the shedding frequency of Strouhal number $St = fD_h/\overline{U_\text{e}} = 3.28$ as well as its harmonic. 
This is the same tone captured numerically by \cite{stack2019turbulence} (33~kHz) and experimentally by \cite{berry2016investigating} (34~kHz) of approximately $St = 3.3$.
This resonant tone has been found to dominate the flow field as well as the experimental far-field acoustics (\cite{berry2016investigating}). Further downstream, the FF0 probe $(x,y)/W = (4.700, 1.300)$ is placed near the nozzle lip and is shown to capture the same resonant tone. The FF0 probe will be used for the remainder of this paper to allow comparison between various control cases, as the M1 probe proves unreliable when actuation is introduced in the $\text{SP}_\text{TE}$ location (see figure~\ref{fig:afc_config}) because it is dominated by the signal of the steady micro-jet.

\begin{figure}
    \centering
    \includegraphics[width=1\textwidth]{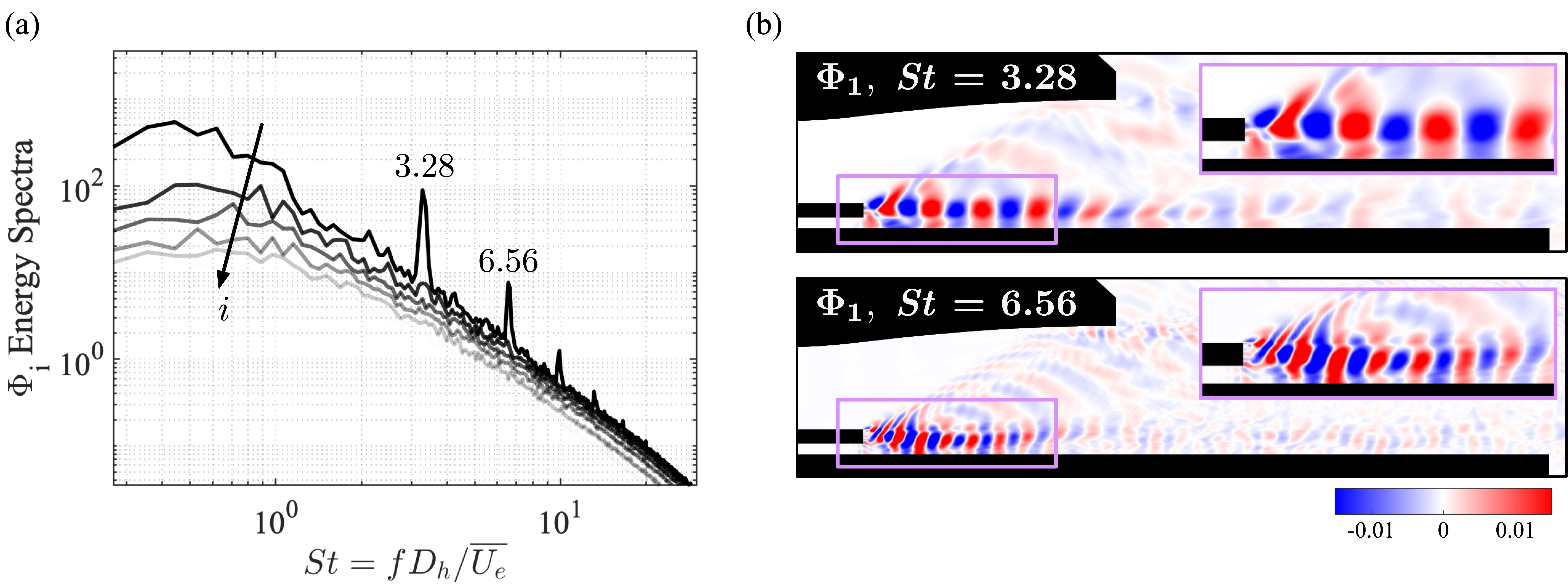}
    \caption{(a) SPOD energy spectrum on the density variable for the baseline flow. (b) Leading modes at representative frequencies.}
    \label{fig:base_spod}
\end{figure}

To gain further insight into the base flow, SPOD is applied to the density variable to extract coherent structures from the flow field, where the identity weight matrix directly measures the variance of the data. Figure~\ref{fig:base_spod}(a) shows the energy distribution for the baseline flow where $\Phi _{i = 1}$ is the leading mode containing the most energy with increasing $i$ indicated with an arrow. Figure~\ref{fig:base_spod}(b) shows the leading mode shapes corresponding to prominent peaks in the energy spectra. Peaks can be seen in the energy spectrum at the dominant frequency of $St = 3.28$ and its harmonics, similar to the M1 pressure spectra shown in figure \ref{fig:base_psd}. Visualization of the leading mode at these frequencies reveals mode structures in the splitter plate shear layer, which further reinforces the dominance of the vortex shedding in the overall flow. Although these modes visualize the density variation, the shape of the modal structures indicates the shedding vortical structures induce pressure waves that leave the nozzle exit. 

\subsubsection{Stability and resolvent analysis}
\begin{figure}
    \centering
    \includegraphics[width=0.9\textwidth]{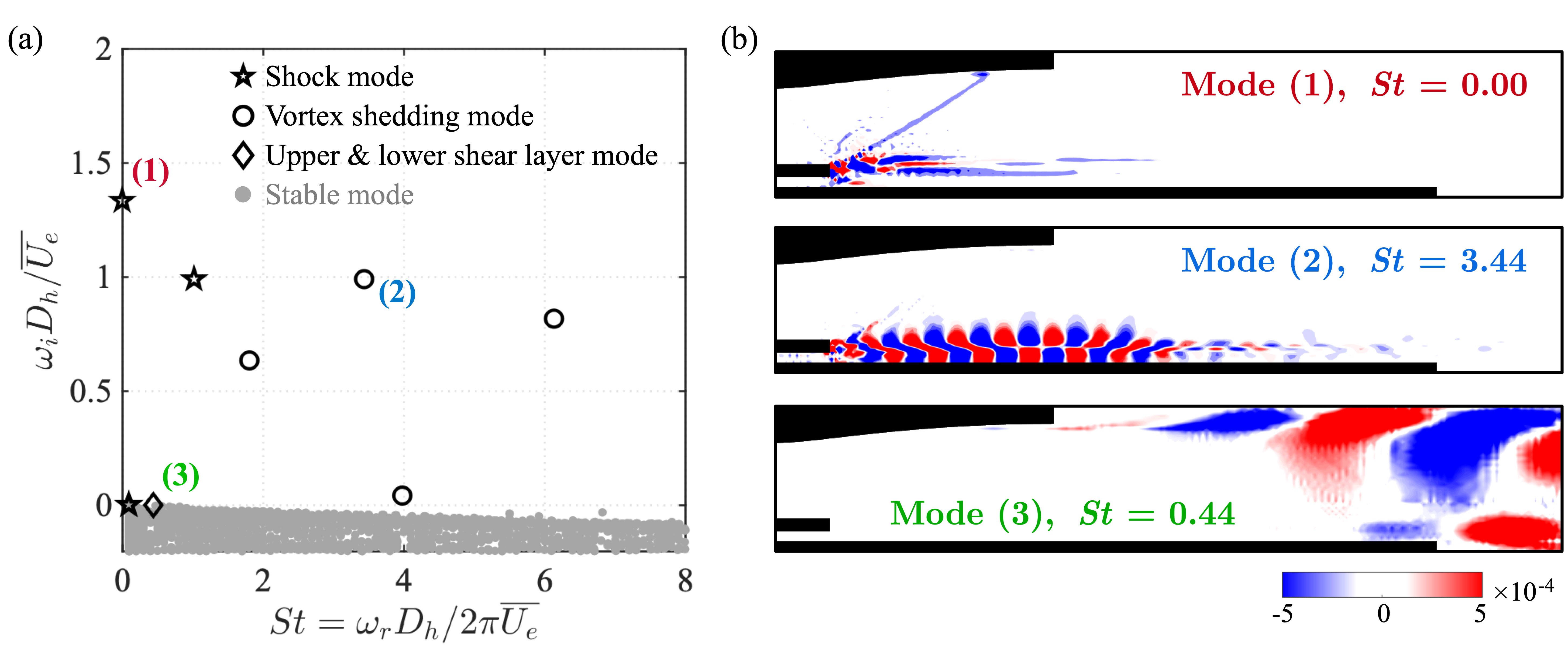}
    \caption{(a) Eigenspectrum for baseline flow. (b) Representative unstable modes for $\hat{u}$.}
    \label{fig:base_stab}
\end{figure}

Stability analysis is performed to identify the dominant instability for the baseline flow, and also to determine an appropriate discount parameter $\alpha$ for the discounted resolvent analysis. Figure~\ref{fig:base_stab}(a) shows the eigenspectrum of the baseline flow with the spanwise wavenumber $\beta = 0$, where eigenvalues corresponding to shock-pattern modes, vortex-shedding modes, and upper and lower shear-layer modes are denoted with different symbols, while all stable modes are in gray. Representative unstable modes in each category with the highest growth rates are shown in figure~\ref{fig:base_stab}(b). It is observed that instabilities corresponding to shock patterns (\includegraphics[width=0.02\textwidth]{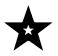}) occur at low frequencies, and the leading eigenmode is a stationary shock mode with $St = 0.00$. The vortex shedding instability (\includegraphics[width=0.02\textwidth]{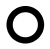}) dominates the mid-to high-frequency range, and the leading vortex shedding eigenmode occurs near the resonant frequency of $St = 3.28$. There is additionally an upper and a lower shear layer instability (\includegraphics[width=0.02\textwidth]{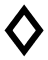}), however, the associated growth rate is relatively small. Based on these results of unstable modes, the discount parameter is chosen to be $\alpha D_h/\overline{U_e} = 3.0$. This value ensures that all unstable eigenvalues in the baseline and control cases (discussed later) will be enclosed.
Generally, choosing a larger discount parameter will result in a decrease in the gain magnitude, however, the mode shapes are mostly unaffected. Prior studies using the discounted resolvent analysis on a laminar airfoil separation (\cite{yeh2019resolvent}) and a turbulent cavity (\cite{liu2021unsteady}) have shown that introducing discounting at larger values removes local spikes in the distribution attributable to subdominant and spurious eigenmodes near the neutral axis, while preserving the overall profile of the gain distribution.

\begin{figure}
    \centering
    \includegraphics[width=0.9\textwidth]{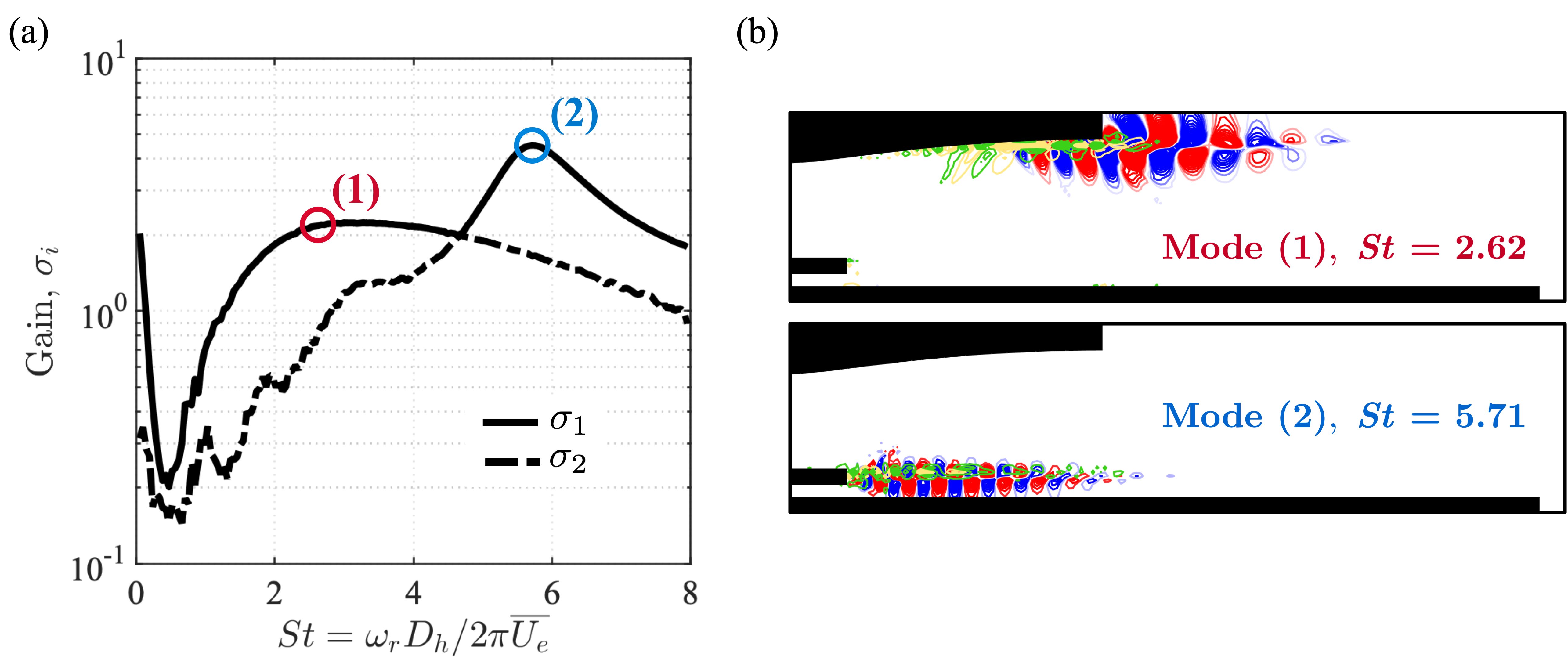}
    \caption{(a) Plot of the gain $\sigma_i$ across a range of frequencies. (b) Optimal forcing and response modes for $\hat{u}$. Yellow-green contour lines represent forcing modes, and blue-red contour lines represent response modes.}
    \label{fig:base_rslvt}
\end{figure}

To uncover how the flow responds to external forcing in a control effort, resolvent analysis is performed with $\beta = 0$ to predict locations that are most receptive to external perturbations. The discounted resolvent operator is used due to the instabilities present in the system, as shown in figure~\ref{fig:base_stab}. Forcing and response mode pairs are obtained at each frequency with a corresponding gain $\sigma_i$. 
Figure~\ref{fig:base_rslvt}(a) shows a plot of the first $\sigma_1$ and second $\sigma_2$ singular values across a range of frequencies. Figure~\ref{fig:base_rslvt}(b) shows the optimal forcing and response mode pairs at representative frequencies, where yellow-green contour lines represent forcing modes and blue-red contour lines represent response modes. In the baseline case, two local peaks are seen with forcing and response modes distributed in the upper shear layer, denoted as Mode~(1), and splitter plate shear layer, denoted as Mode~(2). 

In Mode~(1), the spatial structures of the response mode are observed to originate from inside the nozzle, near the location at which the primary shock impinges on the nozzle wall and induces flow separation. This suggests that the upper shear layer response may be associated with the separation in SR2 (figure~\ref{fig:base_flow}(a)), and reducing the size of the separation can influence the upper shear layer instability. On the other hand, the splitter-plate shear layer forcing-response modes, denoted as Mode~(2), appear at a frequency of $St = 5.71$ rather than the resonant tone. 
This is likely due to the non-normal relation between the two most dominant vortex shedding eigenvalues (see figure~\ref{fig:base_stab}) when performing the discounted resolvent analysis.
In a non-normal system matrix, transient growth may be observed even if all eigenvalues lie in the stable plane, i.e., energy amplification is possible prior to an exponential decay as time approaches infinity (\cite{schmid2014analysis}). Since the eigenvalues inherently describe the time-asymptotic behavior for non-normal systems, they therefore fail to capture processes that occur over finite time scales. In the context of the discounted resolvent, which focuses on short-term dynamics, the identified optimal response can thus occur at a forcing frequency that is far from an eigenvalue, known as pseudoresonance (\cite{trefethen1993hydrodynamic}). \cite{schmid2014analysis} have suggested that the difference between the resolvent norm and the eigenvalue-based response must be attributed to the nonorthogonality of the eigenvectors.

Since the shedding instability is most prevalent in the flow and is the source of the primary resonant tone, the spatial location highlighted by Mode~(2) is utilized to inform the actuator placement rather than Mode~(1). 
The Mode~(2) forcing is observed near the SPTE, indicating that introducing perturbations in this region has the most potential for influencing the development of the shear layer instability. 
Moreover, the leading gain $\sigma_1$ associated with Mode~(2) is greater than Mode~(1), suggesting that acting on Mode~(2) may have a greater capacity of altering the mean flow. This motivates the choice of spatial locations for active flow control previously presented in figure~\ref{fig:afc_config}, where actuators are placed in various locations around this region.

It is also observed that when considering the sub-optimal distribution $\sigma_2$, forcing-response mode pairs pertaining to a single flow feature will follow a continuous curve throughout the frequency range. At $St = 4.63$, there is an intersection between two continuous curves, where the upper shear layer response (Mode~(1)) moves into the sub-optimal distribution while the vortex shedding response (Mode (2)) moves into the optimal distribution. 

\subsection{Active Control: Splitter Plate Top Surface Actuation} 
In this section, modifications introduced from the splitter plate top surface steady-blowing micro-jet actuation are discussed, and new micro-jet-induced flow features are analyzed. SPOD and resolvent analysis are applied to each control case to investigate changes in the forcing-response dynamics of the flow. 

\subsubsection{Instantaneous and mean flow features}
\begin{figure}
    \centering
    \includegraphics[width=1\textwidth]{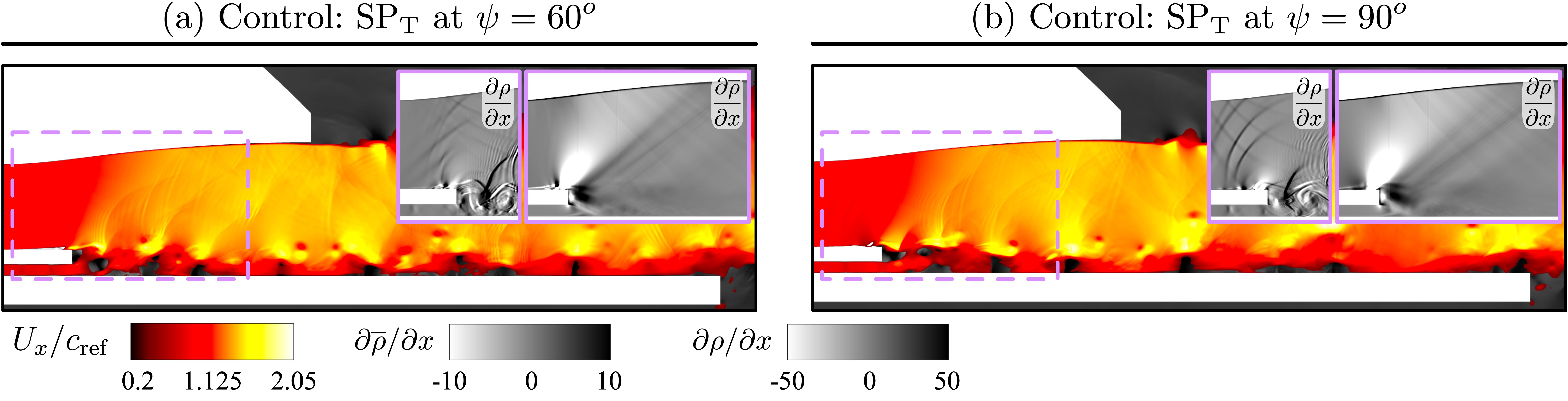}
    \caption{Instantaneous and mean flow fields for control cases with micro-actuation introduced at the $\text{SP}_\text{T}$ location with (a) $\psi = 60\degree$ and (b) $\psi = 90\degree$}
    \label{fig:spt_flow1}
\end{figure}

When steady-blowing is introduced along the top surface of the splitter plate ($\text{SP}_\text{T}$ in figure \ref{fig:afc_config}) at high actuation angles $\psi \in [60\degree, 90\degree]$, the primary shock becomes visibly absent, and two unsteady sets of pressure waves emerge from the micro-jet, as shown in figure~\ref{fig:spt_flow1}. The instantaneous density gradient shows that actuating at $\psi = 90\degree$ causes the unsteadiness to propagate further upstream into the nozzle, and as discussed later, the aft-deck surface loading is also increased the most. Due to this, the flow structures associated with $\psi = 60\degree$ actuation are chosen for further analysis.

\begin{figure}
    \centering
    \includegraphics[width=0.85\textwidth]{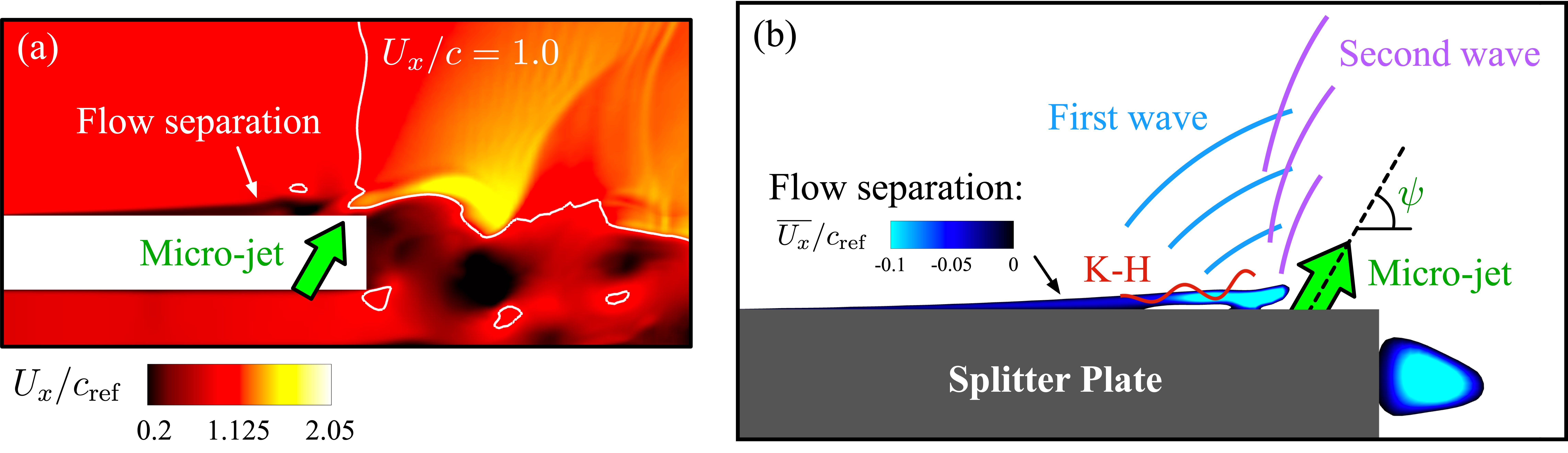}
    \caption{(a) Close-up view with iso-lines of Mach 1.0 and (b) schematic of micro-jet proximity region for $\text{SP}_\text{T}$ at $\psi = 60\degree$.}
    \label{fig:spt_schem}
\end{figure}

Figure~\ref{fig:spt_schem}(a) shows a closer view of the micro-jet at $\psi = 60\degree$ with iso-lines of Mach 1.0 and figure~\ref{fig:spt_schem}(b) illustrates the formation mechanism of the two sets of pressure waves.  The introduction of the micro-jet causes separation upstream of the actuator; the associated shear layer introduces a Kelvin--Helmholtz instability that generates the first set of pressure waves as the instability grows. As the core flow is still expanding above the splitter plate and has yet to reach Mach 1.0, these waves are allowed to propagate upstream. When the unsteady instability reaches and interacts with the micro-jet, a second set of pressure waves, inclined at a higher angle relative to the first, is formed due to the flow-induced flapping motion of the actuation. The two sets of waves propagate upwards, and impinge and reflect off the SERN wall. This interaction between the main stream and the micro-jet causes the incoming main flow to be slightly unsteady as it mixes with the bypass stream, which inhibits the stacking of the compression waves induced by the shedding vortices after the splitter plate. Hence, the S1 shock no longer forms, which additionally eliminates the shock-induced separation region SR2 (figure~\ref{fig:base_flow}). 

\begin{figure}
    \centering
    \includegraphics[width=1\textwidth]{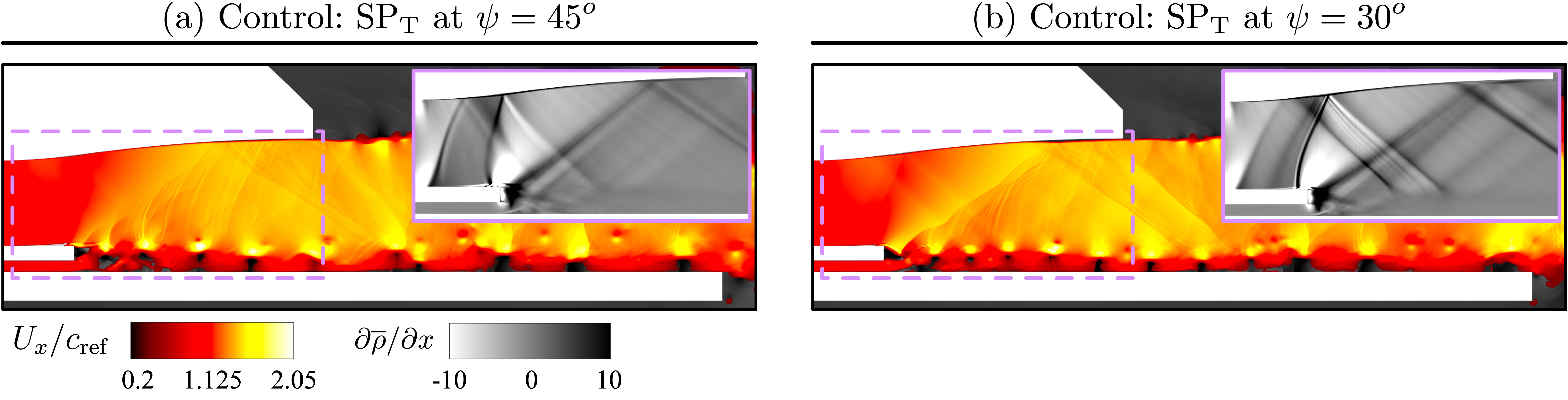}
    \caption{Instantaneous and mean flow fields for control cases with micro-actuation introduced at the $\text{SP}_\text{T}$ location with (a) $\psi = 45\degree$ and (b) $\psi = 30\degree$}
    \label{fig:spt_flow2}
\end{figure}

At low actuation angles $\psi \in [30\degree, 45\degree]$, shown in figure~\ref{fig:spt_flow2}, the separation upstream of the micro-jet is reduced, so that the Kelvin-Helmholtz instability is no longer present. When the main stream encounters the micro-jet, a new relatively steady and stronger shock is formed instead. At $\psi = 45\degree$, this shock is weaker and inclined at a higher angle relative to the shock formed at $30\degree$. In both these low actuation angle cases, the primary shock remains mostly unaffected and thus has minimal control authority over the shock train development.

\subsubsection{Spectral analysis}
\begin{figure}
    \centering
    \includegraphics[width=1\textwidth]{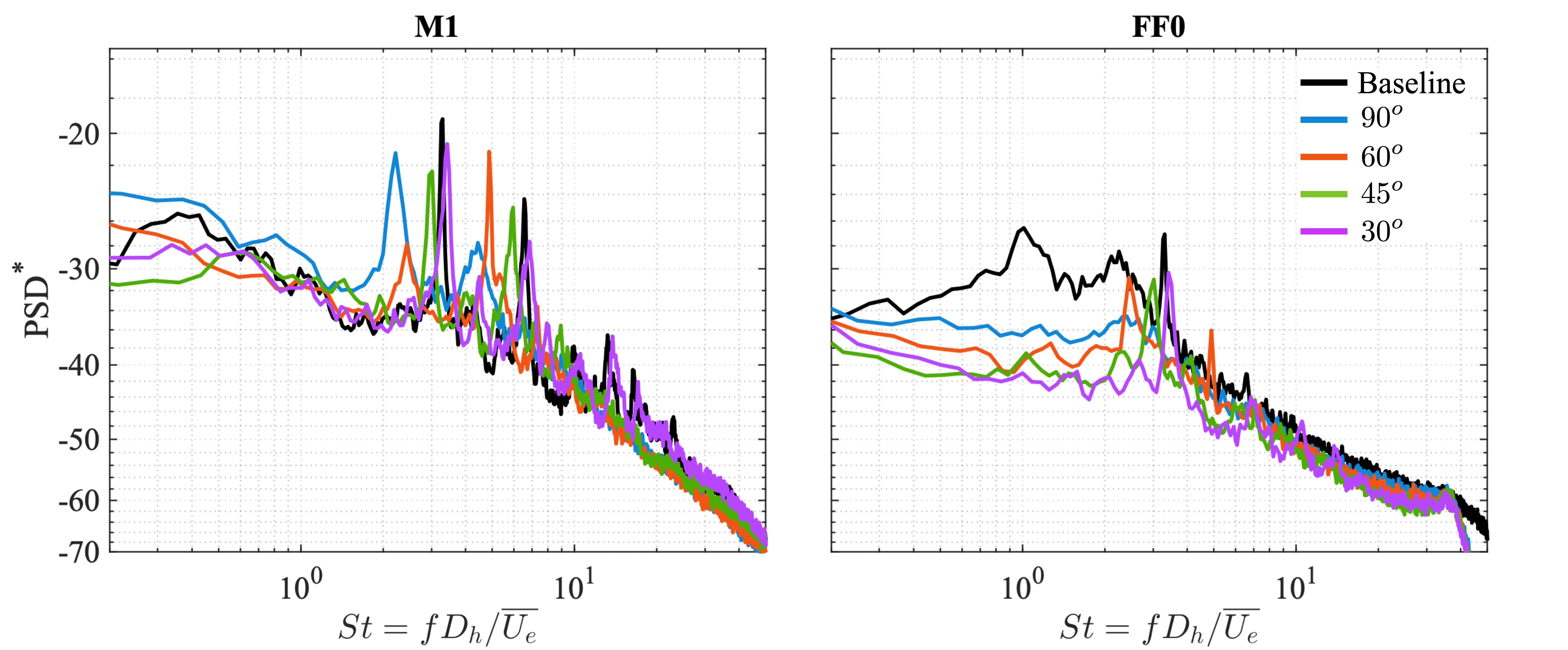}
    \caption{PSD of the pressure time series through the M1 and FF0 probes for all $\text{SP}_{\text{T}}$ control cases.}
    \label{fig:spt_FF0-M1}
\end{figure}

The pressure spectra through the M1 and FF0 probes are shown for all actuation angles in figure \ref{fig:spt_FF0-M1}. It is observed through the M1 location that the shedding frequency decreases at higher actuation angles, and at $60\degree$, the amplitude of the resonant tone is significantly reduced. Meanwhile further downstream, the FF0 location demonstrates the overall broadband content decreases with the angle. As high-angled actuation was found to be more beneficial towards the mitigation of the primary shock development, the most optimal angle at the $\text{SP}_\text{T}$ location is chosen to be $\psi = 60\degree$.

\begin{figure}
    \centering
    \includegraphics[width=1\textwidth]{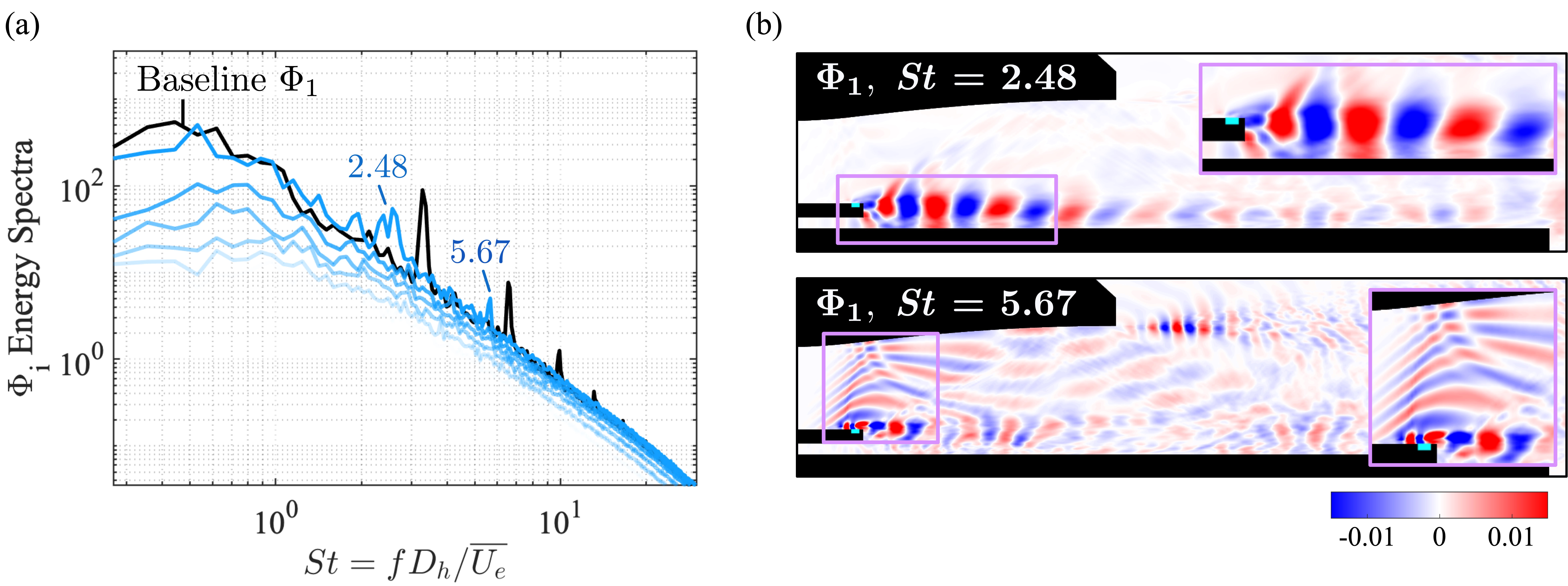}
    \caption{(a) SPOD energy spectrum on the density variable for the $\text{SP}_\text{T}$ optimal control case at $\psi = 60\degree$. Leading baseline energy distribution illustrated for comparison. (b) Leading modes $\Phi_1$ at the frequencies $St = 2.48$ and $St = 5.67$.}
    \label{fig:spt60_spod}
\end{figure}

To further understand how the flow physics are modified, SPOD is performed on the density variable for the optimal case of $\psi = 60\degree$. Figure~\ref{fig:spt60_spod}(a) shows the SPOD energy spectra with the leading baseline distribution illustrated in black for comparison, and corresponding leading modes $\Phi_1$ at representative frequencies are displayed in figure~\ref{fig:spt60_spod}(b). The resonant tone corresponding to the vortex-shedding instability is shifted to a lower frequency of $St = 2.48$ with a reduced amplitude, as was observed in the M1 pressure spectra. This suggests a reduction in the energy content associated with the shedding instability when control is introduced in this configuration. An additional high-frequency peak at $St = 5.67$ is observed, where this frequency is associated with pressure waves originating from the micro-jet that propagate and reflect off the top surface of the SERN. Compared to the baseline flow, where the vortex shedding held the majority of the energy content, the $\text{SP}_\text{T}~60\degree$ case disperses this energy into the unsteadiness generated from the actuator. 

\begin{figure}
    \centering
    \includegraphics[width=0.9\textwidth]{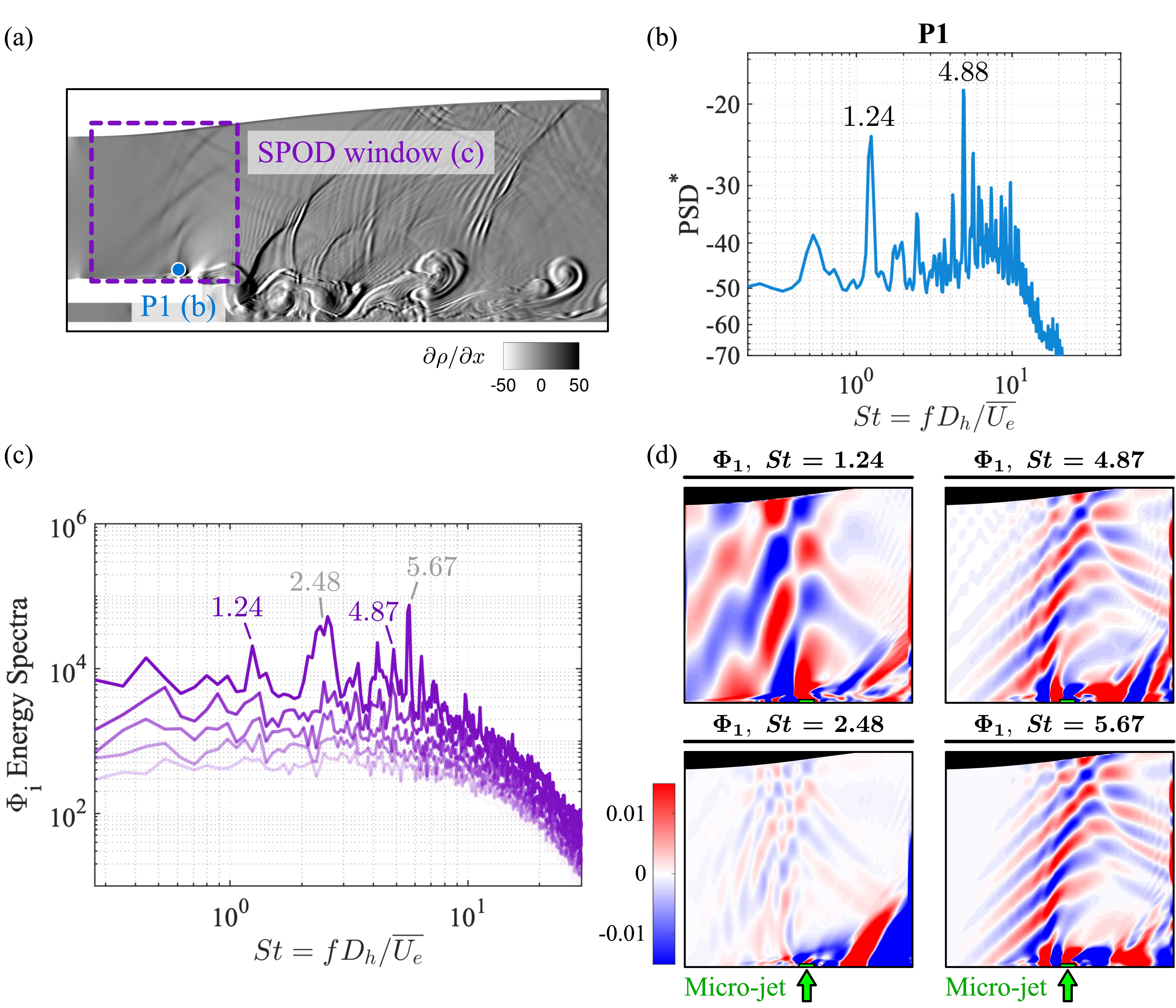}
    \caption{Spectral analysis for the $\text{SP}_\text{T}$ control case at $\psi = 60\degree$ near the micro-jet proximity region. (a) Location of P1 probe and reduced spatial window for SPOD. (b) PSD of the pressure time series through P1. (c) SPOD energy spectrum on the stream-wise density gradient variable. (d) Leading modes $\Phi_1$ at representative frequencies.}
    \label{fig:spt60_spod_act}
\end{figure}

Although two separate sets of pressure waves can be distinguished in the instantaneous flow field, as shown earlier in figure~\ref{fig:spt_flow1}(a), only one peak in the SPOD spectrum is found to correspond to the unsteady waves. To identify the frequencies corresponding to the two sets of pressure waves, the PSD of a probe, marked P1, and the SPOD of the stream-wise density gradient is computed in the micro-jet proximity region. Figure~\ref{fig:spt60_spod_act}(a) shows the location of the P1 probe $(x,y)/W = (3.966, 0.980)$ as well as the reduced spatial window used for SPOD, where P1 is placed directly behind the micro-jet within the unsteadiness to capture their signals. The results are then re-interpolated onto a $150 \times 100$ uniform grid within the reduced SPOD window to resolve the two sets of pressure waves. SPOD of the instantaneous density field did not provide a clear delineation of the two sets of pressure waves of interest; however, the gradient of the density field successfully highlighted the mechanism of interest, as shown in figure~\ref{fig:spt_flow1}, and is used instead.

Figure~\ref{fig:spt60_spod_act}(b) shows the pressure spectra through P1, where two frequencies $St = 1.24$ and $St = 4.88$ are captured. The SPOD is used as a tool to visualize the spatial structures at those frequencies, shown in figure~\ref{fig:spt60_spod_act}(c-d). In the SPOD energy spectrum, the same frequencies as those captured in the pressure spectra are found, where the corresponding leading modes distinctly illustrate the two unsteady waves. The spatial structures of their modes indicate that the higher frequency $St = 4.87$ corresponds to the first set of waves, while the lower frequency $St = 1.24$ corresponds to the second set of waves. It is also noted that despite using a reduced spatial window, the frequencies $St = 2.48$ and $St = 5.67$ captured in the larger domain SPOD remain dominant in these energy spectra. Visualizing their modes additionally shows similar structures presented earlier in figure~\ref{fig:spt60_spod}. Although the frequencies $St = 1.24$ and $St = 4.87$ directly correspond to the unsteady pressure waves, their combined influence appears through the $St = 5.67$ tone.

\subsubsection{Instabilities and resolvent modes}
\begin{figure}
    \centering
    \includegraphics[width=1\textwidth]{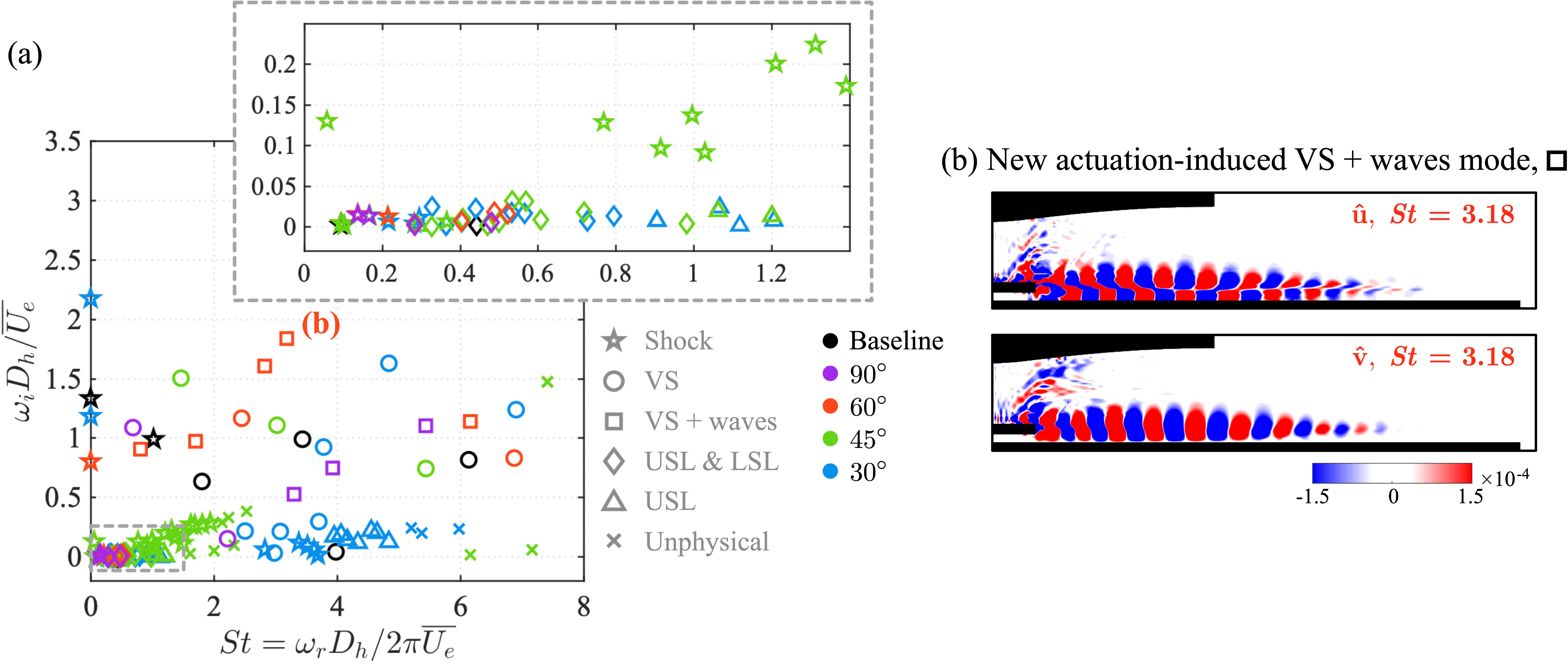}
    \caption{(a) Eigenspectrum for all control cases at the $\text{SP}_\text{T}$ location with a zoomed-in view shown in a dashed gray box. Stable modes are not plotted for clarity. (b) New actuation-induced unstable mode.}
    \label{fig:spt_stab}
\end{figure}

Stability analysis is performed for all angles $\psi$ considered at the $\text{SP}_\text{T}$ location, where the linearized operator $\mathcal{L}_{\overline{\textbf{q}}}$ is reconstructed using the corresponding time-averaged flow obtained from each control case. Figure~\ref{fig:spt_stab} shows the eigenspectrum for all $\text{SP}_\text{T}$ control cases, where only the unstable eigenvalues ($\omega_i > 0$) are shown. At high angles $\psi \in [60\degree, 90\degree]$, a new actuation-induced instability with a relatively high growth rate is found, denoted with (\includegraphics[width=0.02\textwidth]{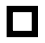}). In this mode, the unsteadiness generated from the actuator can be observed, and the spatial structures of the vortex shedding (VS) instability develop from the micro-jet rather than the splitter plate trailing edge, as previously shown in figure~\ref{fig:base_stab}(b). At $\psi = 90\degree$, the new unsteady mode is found predominantly in the mid-to high-frequency range ($St \in [3.30, 5.43])$, while decreasing the angle to $\psi = 60\degree$ shifts this mode into a wider range of frequencies ($St \in [0.81, 6.16])$. 
Although these eigenmodes occur with high growth rates compared to the baseline spectrum, the unsteadiness associated with micro-jet-induced pressure waves proves beneficial in eliminating the primary shock, as shown in figure~\ref{fig:spt_flow1}. It is also noted that the new actuation-induced mode is more unstable at $\psi = 60\degree$ compared to $\psi = 90\degree$, suggesting this unstable behavior prompted the $\psi = 60\degree$ case to perform better than the $\psi = 90\degree$ case. 

Additional shock and upper and lower shear layer instabilities are also observed at high angles in figure~\ref{fig:spt_stab}(a); however, both their frequencies and growth rates are relatively small. As the actuation angle decreases to $\psi \in [30\degree, 45\degree]$, the eigenspectrum becomes largely concentrated with shock and shear layer modes, where these instabilities are mostly found in the low-frequency range for $\psi = 45\degree$ and in the mid-frequency range for $\psi = 30\degree$. The most unstable eigenmode across all control cases occurs at $\psi = 30\degree$ and corresponds to a stationary shock instability, indicating that introducing actuation closer along the incoming flow direction amplifies the shock unsteadiness behavior.

\begin{figure}
    \centering
    \includegraphics[width=0.8\textwidth]{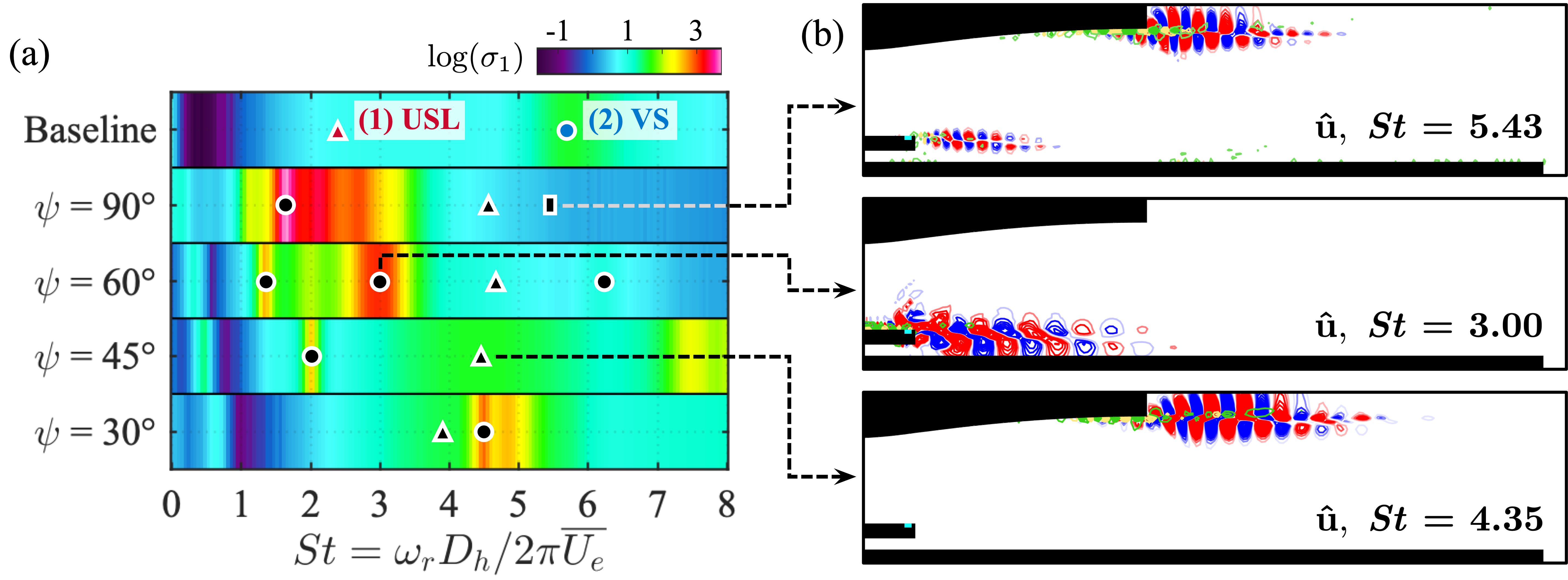}
    \caption{(a) Distribution of the optimal resolvent gain $\sigma_1$ across a range of frequencies for all $\text{SP}_\text{T}$ cases. Symbols represent different types of forcing-response mode pairs. (b) Representative forcing and response modes with optimal gain for $\hat{u}$. Forcing modes colored as yellow-green and response modes colored as blue-red.}
    \label{fig:spt_rslvt}
\end{figure}

Discounted resolvent analysis is performed with $\alpha D_h/\overline{U_e} = 3.0$ for all control cases to assess how the forcing-response dynamics of the flow have been changed after actuation is introduced. Figure~\ref{fig:spt_rslvt}(a) shows the distribution of the leading gain $\sigma_1$ for all angles $\psi$, and is presented as a contour plot across a range of frequencies. In the baseline distribution, the location of the upper shear layer and vortex shedding forcing-response mode pairs, previously shown in figure~\ref{fig:base_rslvt}, are denoted as (\includegraphics[width=0.02\textwidth]{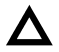}) and (\includegraphics[width=0.02\textwidth]{2D_jet/figures/shp_circle.png}), respectively. For each case, forcing and response modes are obtained and visualized at local frequency peaks and throughout the distribution. Figure~\ref{fig:spt_rslvt}(b) presents representative forcing (yellow-green contour lines) and response (blue-red contour lines) modes, where $\hat{u}$ is plotted. 

When control is introduced at $\psi = 90\degree$, the VS mode is shifted to a lower frequency with a higher gain while the USL mode is shifted to a higher frequency with a low gain. At higher frequencies, a new forcing-response pair (\includegraphics[width=0.02\textwidth]{2D_jet/figures/shp_square.png}) appears with a lower energy amplification, with the spatial structure of this response mode evident in both the upper and splitter plate shear layers. The USL response appears at the highest frequency when $\psi = 60\degree$, and the VS response can additionally be found at multiple frequencies. As $\psi$ decreases, the USL response moves towards a lower frequency while the VS response shifts to a higher frequency; at $\psi = 30\degree$, the two types of forcing-response modes appear most similar to the baseline case. Introducing control at the splitter plate top surface generally increases the responsive frequency of the USL, while simultaneously lowering the frequency of the VS response. Further, the USL resolvent modes always appear as a broadband peak in the gain distribution, while VS modes appear as a sharp peak and can be distributed into multiple frequencies. Introducing control shifts the frequency at which these modes appear; with the specific values being sensitive to the angle of actuation.

Contrary to Mode~(1) shown previously in figure~\ref{fig:base_rslvt}, the spatial structures of all USL response modes in the control cases originate closer to the nozzle lip rather than from inside the nozzle, and the structures of the corresponding forcing modes are smaller. This suggests that the USL resolvent modes are spatially associated with the size of flow separation in SR2, as all $\text{SP}_\text{T}$ control cases either reduce or eliminate the size of separation (see figures~\ref{fig:spt_flow1}, \ref{fig:spt_flow2}). Similarly, the spatial structures of all VS response modes in the controlled flows are initiated from the top of the splitter plate surface, indicating this control location is influencing the spatial formation of the signature shedding instability such that it develops closer to the micro-jet. Additionally, as the actuation angle $\psi$ increases, the corresponding forcing modes become purely concentrated along the top surface of the splitter plate upstream of the actuator, rather than the SPTE proximity region as in the baseline case.

\subsection{Active Control: Splitter Plate Trailing Edge Surface Actuation} 
This section discusses the effects of actuation at the splitter plate trailing edge surface. In contrast to the splitter plate top surface actuation, this control location directly influences the size of the shedding vortices. SPOD is applied to the most optimal case, while resolvent analysis is performed on all control cases. 

\subsubsection{Instantaneous and mean flow features}
\begin{figure}
    \centering
    \includegraphics[width=0.8\textwidth]{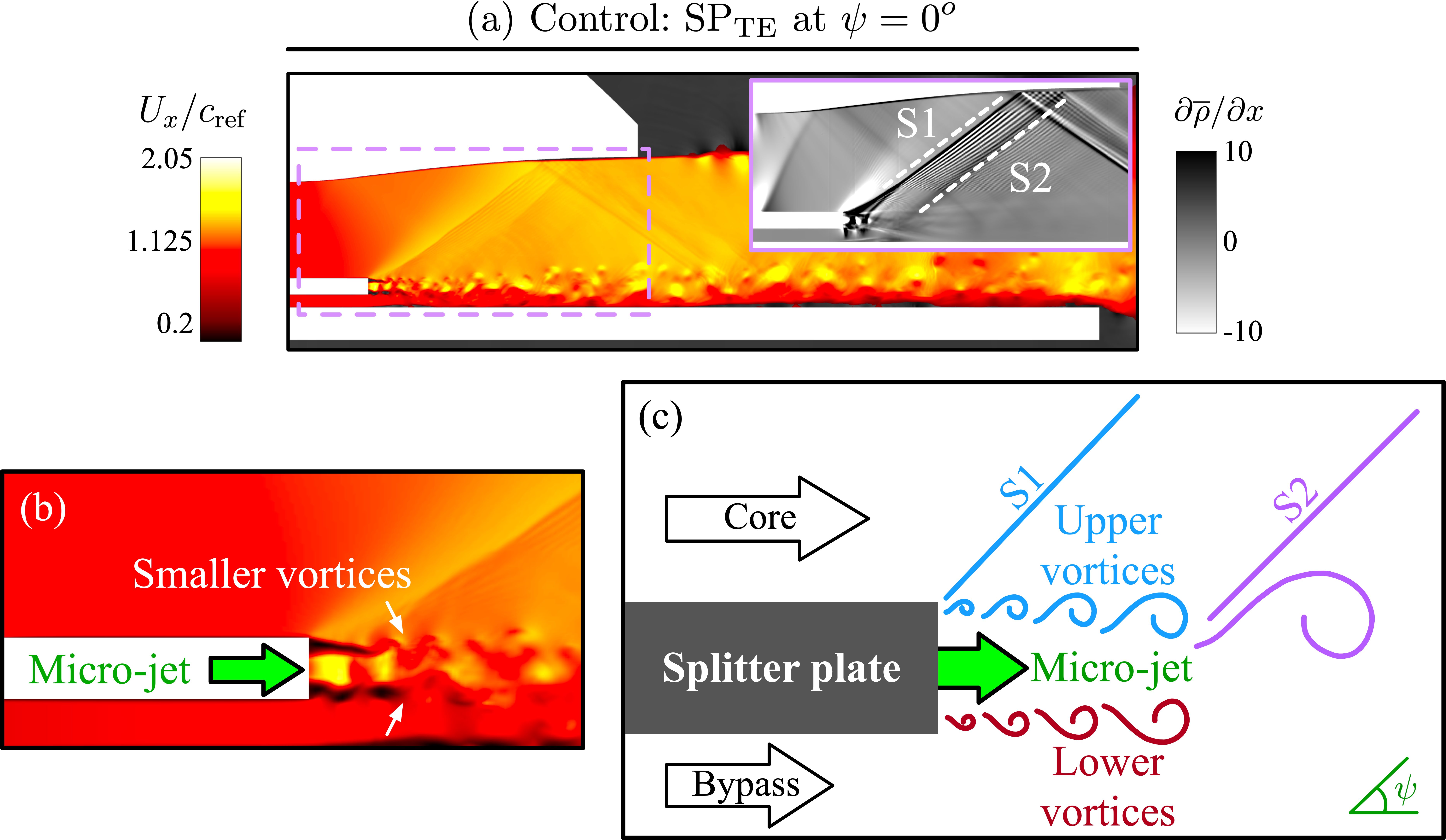}
    \caption{(a) Instantaneous and mean flow fields for control cases at the $\text{SP}_\text{TE}$ location with $\psi = 0\degree$. (b) Closer view and (c) schematic of micro-jet proximity region (not to scale).}
    \label{fig:spte_flow1}
\end{figure}

\begin{figure}
    \centering
    \includegraphics[width=1\textwidth]{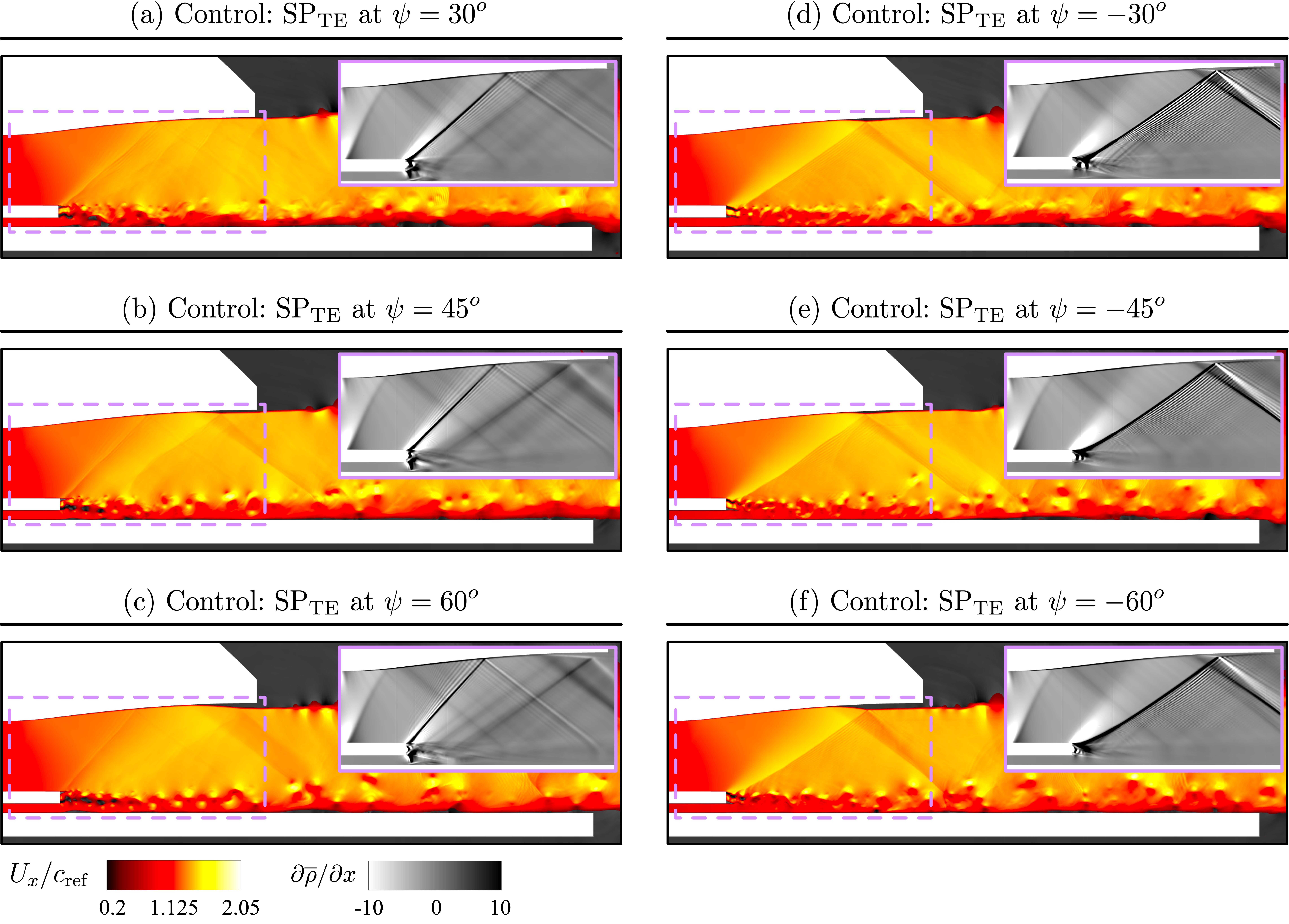}
    \caption{(a) Instantaneous and mean flow fields for control cases at the $\text{SP}_\text{TE}$ location with (a-c) $\psi > 0\degree$ and (d-f) $\psi < 0\degree$.}
    \label{fig:spte_flow2}
\end{figure}

When steady-blowing is introduced at the trailing edge surface of the splitter plate ($\text{SP}_\text{TE}$ in figure \ref{fig:afc_config}), the micro-jet directly interferes with the mixing of the main and bypass streams and considerably smaller vortices are generated for all angles $\psi$ considered, as shown in figures~\ref{fig:spte_flow1}(a) and \ref{fig:spte_flow2}. When the micro-jet is introduced in the same direction as the incoming main and bypass streams ($\psi = 0\degree)$, shown in figure~\ref{fig:spte_flow1}(a), a secondary shock S2 is formed. Figure~\ref{fig:spte_flow1}(b) shows a close-up view of the micro-jet at $\psi = 0\degree$, and figure~\ref{fig:spte_flow1}(c) illustrates the formation mechanism of the secondary shock. The micro-jet breaks down the original thick splitter plate shear layer into two thinner shear layers. As a result, upper and lower sets of vortical structures are generated, whose sizes are smaller compared to the vortices in the baseline flow (figure~\ref{fig:base_flow}). Moreover, the size of the upper and lower sets of vortices relative to each other are disparate due to the difference in the local speed of the main and bypass streams when mixing with the micro-jet. Further downstream of the SPTE, the upper and lower vortices interact and merge, essentially where the potential core due to the micro-jet collapses, forming a secondary shock S2. 

When $\psi$ is increased in the $30\degree$ to $60\degree$ range, shown in figure~\ref{fig:spte_flow2}~(a-c), the upper vortices become smaller, reducing the strength of the resulting compression waves and ultimately weakening the S1 shock. Eventually, S1 is formed almost entirely due to the flow deflection caused by the micro-jet, and the contribution from the upper vortex-induced compression waves is negligible. Meanwhile, the S1 shock is inclined at a higher angle relative to the baseline flow due to the micro-jet being angled upwards, and the S1 shock moves further upstream as $\psi$ increases. Additionally, the size of the lower vortices relative to the upper vortices becomes larger, resulting in the strengthening of the S2 shock when the two sets of vortices interact and merge. This causes the separation in SR2 (figure~\ref{fig:base_flow}) to be induced by the S2 shock rather than the S1 shock. At the optimal angle where $\psi = 30\degree$, neither the S1 nor S2 shocks will induce the flow separation in SR2 near the nozzle lip. Simultaneously, the weakened S1 shock leads to a smaller difference in the stream-wise velocity before and after the S1 shock, such that the flow processed by S2 is of higher speed. This results in the impingement of S2 further downstream on the SERN and closer to the nozzle lip. As $\psi$ increases, the S1 and S2 shocks gradually separate and are formed apart from each other. 

At negative angles $\psi < 0\degree$, shown in figure \ref{fig:spte_flow2} (d-f), the upper vortices contribute weaker compression waves to the S1 shock. However, a stronger oblique shock is formed due to the micro-jet deflecting the main flow downwards before aligning with the horizontal direction. This leads to the overall strengthening of the S1 shock where further decreasing $\psi$ will increase the strength of S1 and consequently, the size of the separation in SR2.

\subsubsection{Spectral analysis}
\begin{figure}
    \centering
    \includegraphics[width=0.9\textwidth]{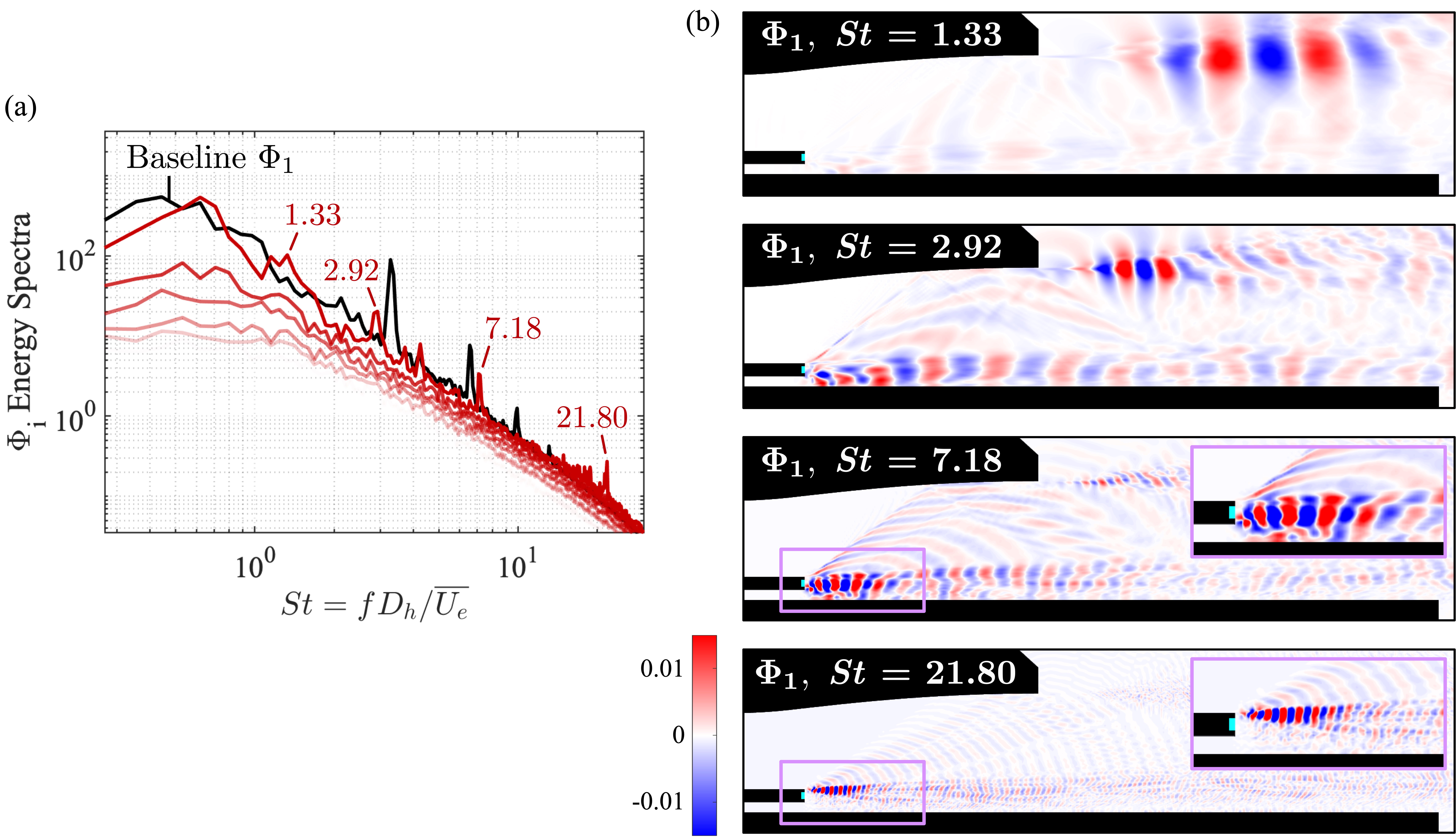}
    \caption{(a) SPOD energy spectrum on the density variable for the $\text{SP}_\text{TE}$ optimal control case at $\psi = 30\degree$. Leading baseline energy distribution illustrated for comparison. (b) Leading modes at representative frequencies.}
    \label{fig:spte30_spod}
\end{figure}

SPOD  energy spectra and modes at representative frequencies are displayed in figure~\ref{fig:spte30_spod} for the optimal control case of $\psi = 30\degree$. Peaks are captured in the leading energy distribution at multiple frequencies with reduced amplitudes compared to the baseline distribution. In the low-frequency range, this control strategy excites frequencies of $St = 1.33$ and $St = 2.92$, corresponding to the USL instabilities. Peaks are observed in the high-frequency range at $St = 7.18$ and $St = 21.80$. Modal structures at the former value show the lower shedding vortices generated from the micro-jet-bypass streams mixing, while the latter value highlights the upper shedding vortices generated from the mixing of the micro-jet and main streams. The higher gradients in the spatial structures of the upper shedding vortices additionally suggest they dissipate faster than the lower vortices. 

\subsubsection{Instabilities and resolvent modes}
\begin{figure}
    \centering
    \includegraphics[width=0.75\textwidth]{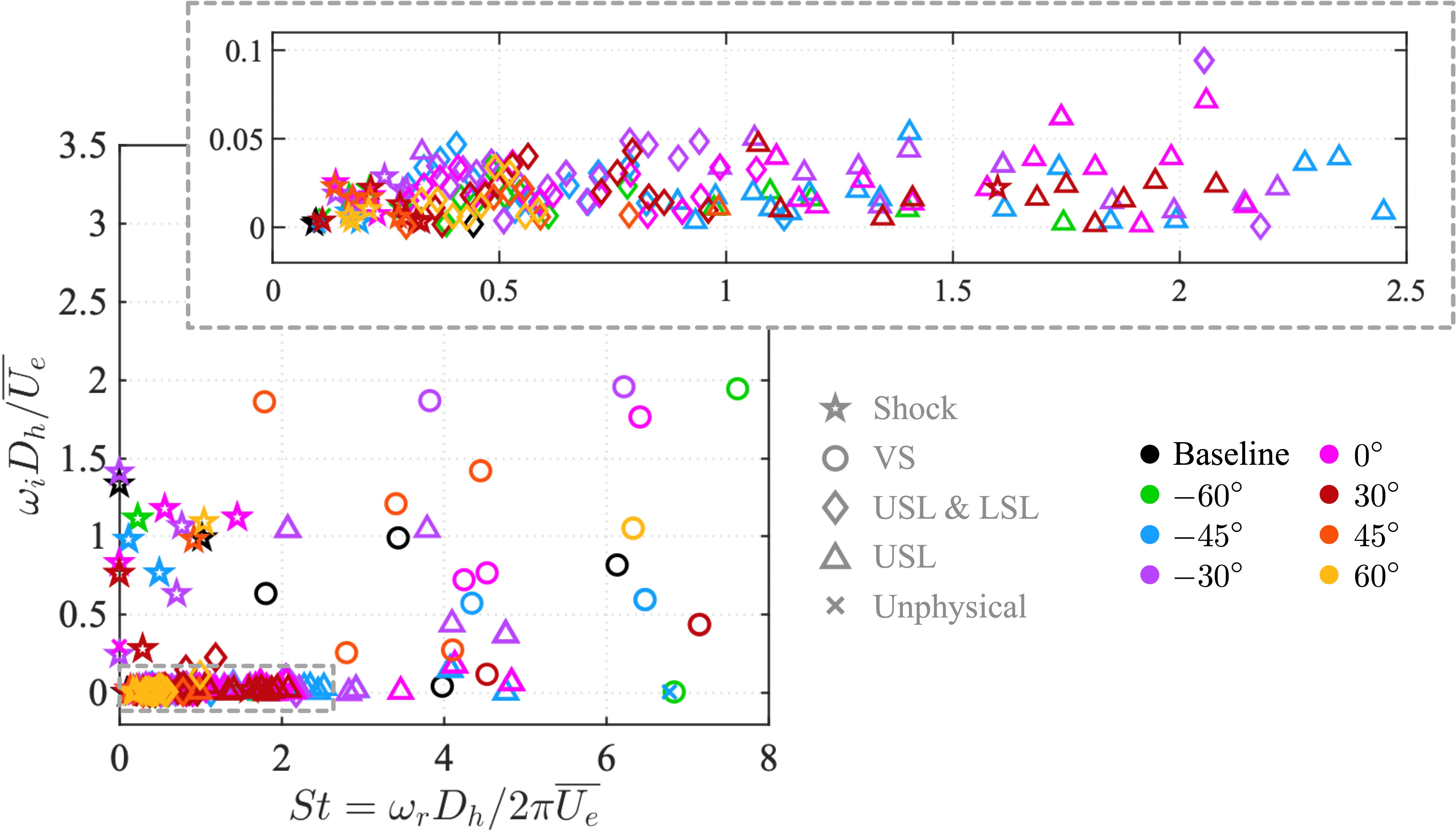}
    \caption{Eigenspectrum for all control cases at the $\text{SP}_\text{TE}$ location with a zoomed-in view shown in a dashed gray box. Stable modes are not plotted for clarity.}
    \label{fig:spte_stab}
\end{figure}

Eigenspectra obtained from the stability analysis of all control cases at the $\text{SP}_\text{TE}$ location are shown in figure~\ref{fig:spte_stab}. Similar to low angle actuation $\psi \in [30\degree, 45\degree]$ at the $\text{SP}_\text{T}$ location, the shock (\includegraphics[width=0.02\textwidth]{2D_jet/figures/shp_star.png}) and upper and lower shear layer instabilities (\includegraphics[width=0.02\textwidth]{2D_jet/figures/shp_diamond.png}) become significantly more unstable compared to the baseline spectrum. This might indicate that introducing actuation with a larger stream-wise component will prompt an amplification in the shock and shear layer behavior through mechanisms that agglomerate the flow deflection and vortex-induced compression waves that synthesize the primary S1 shock. These eigenmodes are distributed across a wider range of frequencies compared to the $\text{SP}_\text{T}$ low-angle actuation cases, shown in figure~\ref{fig:spt_stab}. The zoomed-in view of the dashed box region in figure~\ref{fig:spte_stab} shows that for all angles considered, the upper shear layer instability (\includegraphics[width=0.02\textwidth]{2D_jet/figures/shp_triangleU.png}) spans a large range of frequencies ($St \in [0.0, 2.5]$) while the upper and lower shear layer instabilities (\includegraphics[width=0.02\textwidth]{2D_jet/figures/shp_diamond.png}) appear together in a narrower and lower frequency range ($St \in [0.0, 1.2]$). However, the leading eigenvalue with the lowest growth rate $\omega_i$ across the baseline and all control cases occurs at the optimal angle $\psi = 30\degree$. This might suggest that reducing the size of the shedding vortices and achieving a balance between the primary and secondary shocks can most effectively suppress the baseline instabilities and also stabilize the controlled flow.

\begin{figure}
    \centering
    \includegraphics[width=0.8\textwidth]{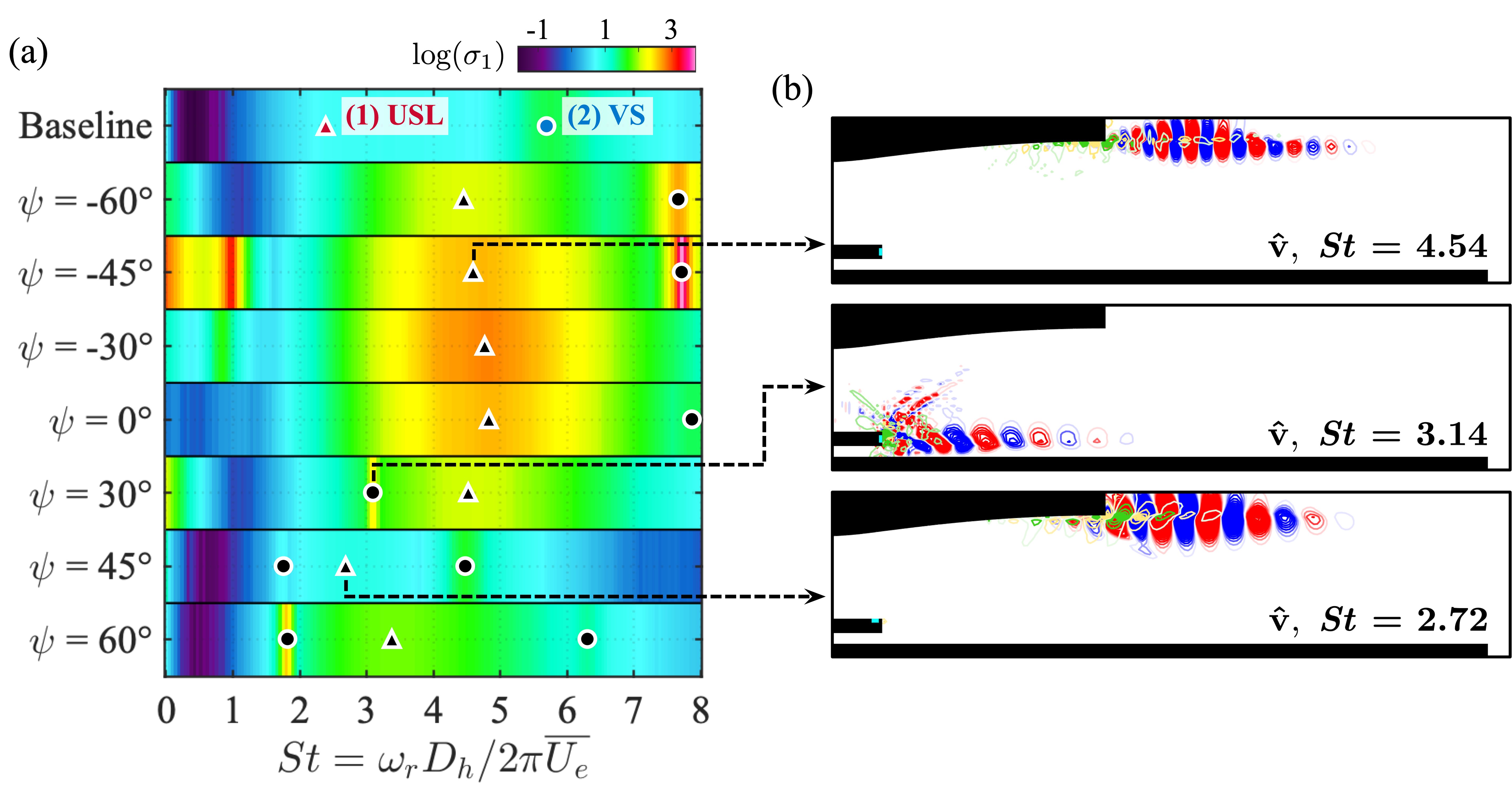}
    \caption{(a) Contours of the optimal resolvent gain $\sigma_1$ across a range of frequencies for all $\text{SP}_\text{TE}$ control cases. (b) Representative forcing-response mode pairs with optimal gain for $\hat{v}$. Forcing modes colored as yellow-green and response modes colored as blue-red.}
    \label{fig:spte_rslvt}
\end{figure}

Resolvent results of all control cases at the $\text{SP}_\text{TE}$ location are shown in figure~\ref{fig:spte_rslvt} in the form of the leading gain $\sigma_1$ for all angles $\psi$ and representative forcing-response mode pairs. 
When actuation is introduced at $\psi \leq 0\degree$, the USL and VS responses are both shifted to a higher frequency. At $\psi = -30\degree$, the VS response is not present in the most optimal gain distribution $\sigma_1$, but rather in the sub-optimal distribution $\sigma_2$. Meanwhile, a ``hotspot'' corresponding to the USL response is evident around $St = 4.5$. This suggests that when actuation is introduced at a negative angle, the USL becomes progressively more responsive to perturbations, as indicated by the larger gain values. Simultaneously, the frequency of the USL response has minimal change across $\psi$, suggesting that although the USL becomes more responsive, it is not sensitive to changes in the actuation angle when $\psi \leq 0\degree$.

When $\psi \geq 0\degree$, the corresponding gain and frequency of the USL response begins to decrease dramatically, implying this region of the flow becomes more stable; however, the USL becomes significantly more sensitive to the changes in $\psi$. The frequency of the VS response decreases as $\psi$ increases and begins to appear at multiple local peaks. This is likely due to the upper and lower shedding vortices (see figure~\ref{fig:spte_flow1}(c)) becoming more apparent in the flow at positive actuation angles, whereas when $\psi < 0\degree$, the lower vortices are inhibited in their development through the core flow (figure \ref{fig:spte_flow2} (d-f)) and the upper shedding vortices are largely dominant. Furthermore, the VS forcing modes concentrate closer to the splitter plate trailing edge, and the corresponding response modes propagate into the primary shock. 

Compared to the baseline case, the USL response in all control cases originates closer to the nozzle lip, similar to the $\text{SP}_\text{T}$ control shown in figure~\ref{fig:spt_rslvt}. The size of the separation region SR2 is either reduced or eliminated in the controlled flows. This further signifies that the spatial response of the USL is largely associated with the SR2 separation. It is also observed that the USL resolvent modes consistently appear as a broadband peak while the VS modes appear as a sharp peak, as was observed in the $\text{SP}_\text{T}$ control.

\subsection{Active Control: Splitter Plate Bottom Surface Actuation} 
\begin{figure}
    \centering
    \includegraphics[width=0.95\textwidth]{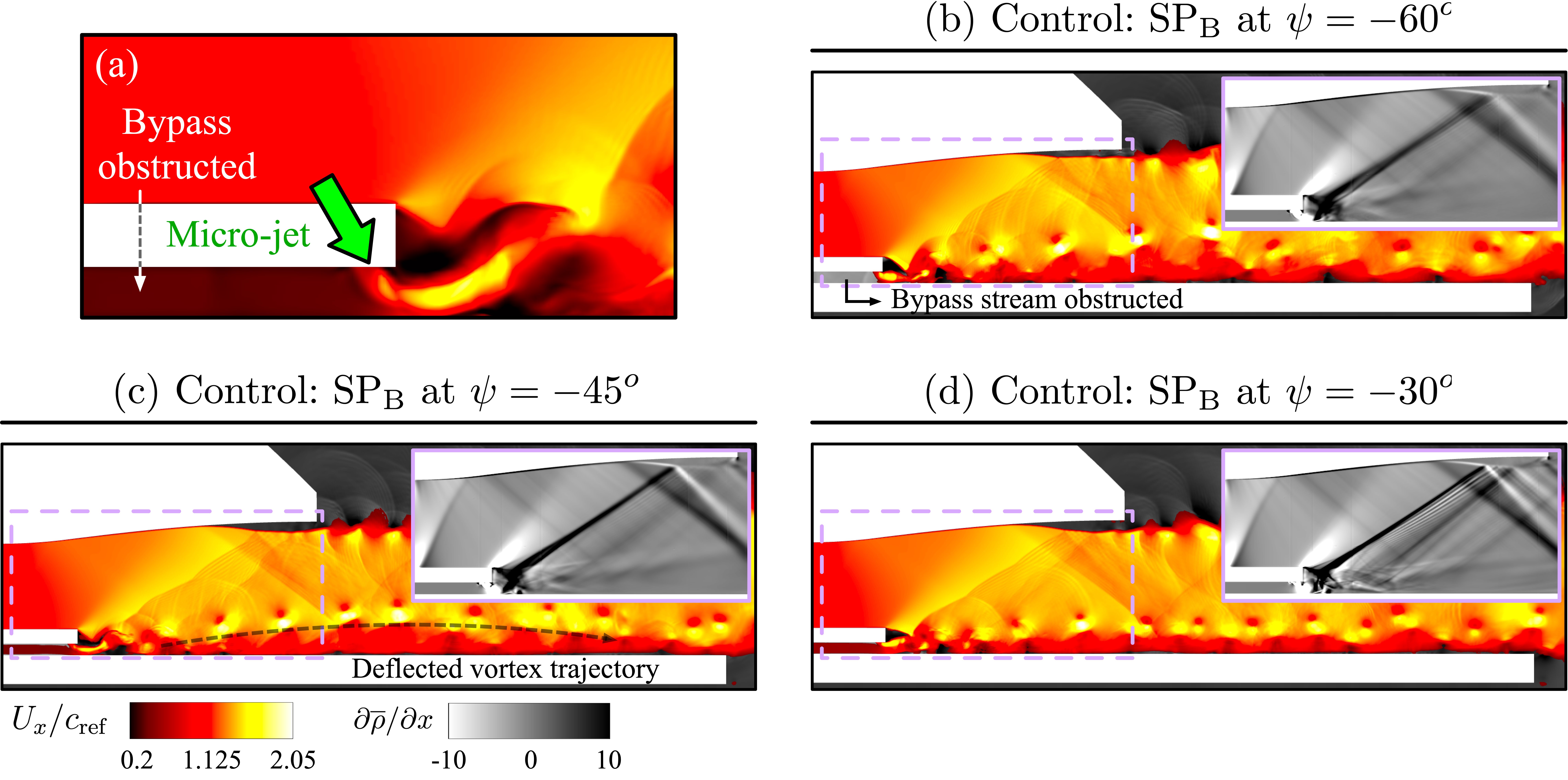}
    \caption{Instantaneous flow fields for control cases at the $\text{SP}_\text{B}$ location. (a) Closer view of micro-jet region, (b) $\psi = -60\degree$, (c) $\psi = -45\degree$, and (d) $\psi = -30\degree$.}
    \label{fig:spb_flow}
\end{figure}

When steady-blowing is introduced at the bottom surface of the splitter plate ($\text{SP}_\text{B}$ in figure \ref{fig:afc_config}), the bypass stream is obstructed for all angles considered and is most apparent in figure \ref{fig:spb_flow} (a-b) when $\psi = -60\degree$. The micro-jet stream reflects from the aft-deck, and the clockwise turn angle of the main stream around the SPTE is larger compared to the baseline flow. The main stream mixes with the micro-jet rather than the bypass stream, causing the flapping motion at the SPTE to be amplified. This generates stronger compression waves and strengthens the S1 shock. The interaction between the actuator and main stream additionally causes both the formation of the shedding vortices and the S1 shock to be spatially delayed such that both develop further downstream from the SPTE compared to the baseline flow (figure~\ref{fig:base_flow}). As a result, the S1 shock impinges on the SERN closer to the nozzle lip and induces a larger separation in SR2. Meanwhile, the deflected micro-jet from the aft-deck diverts the trajectory of the shedding vortices upward, causing the vortices to develop through the core flow where they follow a parabolic-like trajectory indicated by a black dashed line in figure \ref{fig:spb_flow}(c). 

\begin{figure}
    \centering
    \includegraphics[width=0.55\textwidth]{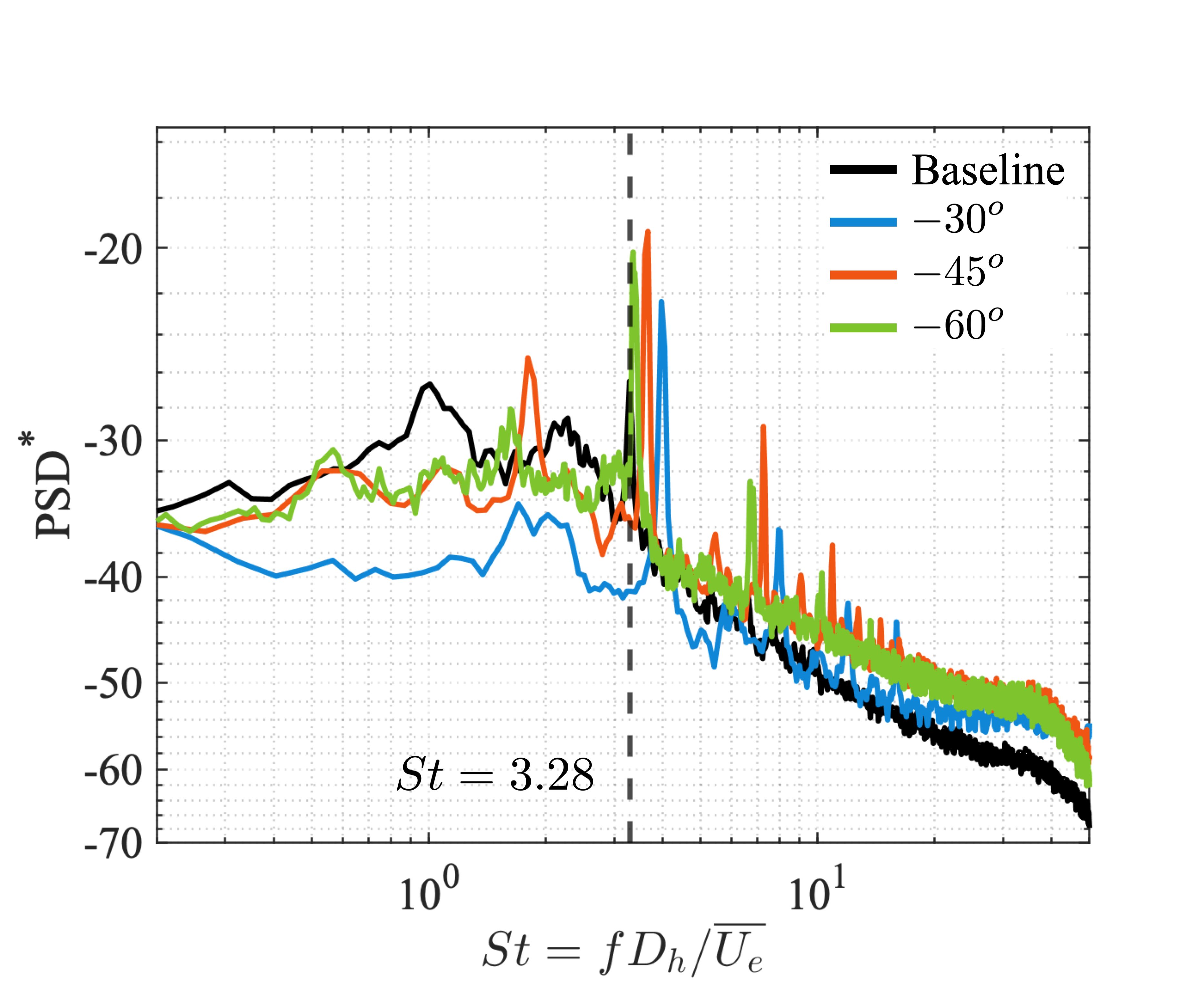}
    \caption{Pressure spectra of the signal from the FF0 point probe for $\text{SP}_\text{B}$ actuation at the angles $-30\degree$, $-45\degree$, and $-60\degree$.}
    \label{fig:spb_psd}
\end{figure}

The PSD at the FF0 probe near the nozzle lip (figure~\ref{fig:spb_psd}) shows that the resonant tone and all harmonics are shifted to a higher frequency. These frequency peaks are significantly amplified for all angles considered at this location. Generally, introducing steady-blowing control at the $\text{SP}_\text{B}$ location is ineffective in achieving the control objectives and will instead enhance the dominant tone and strengthen the primary shock. 

\subsection{Aft-deck surface loading} 
\begin{figure}
    \centering
    \includegraphics[width=0.95\textwidth]{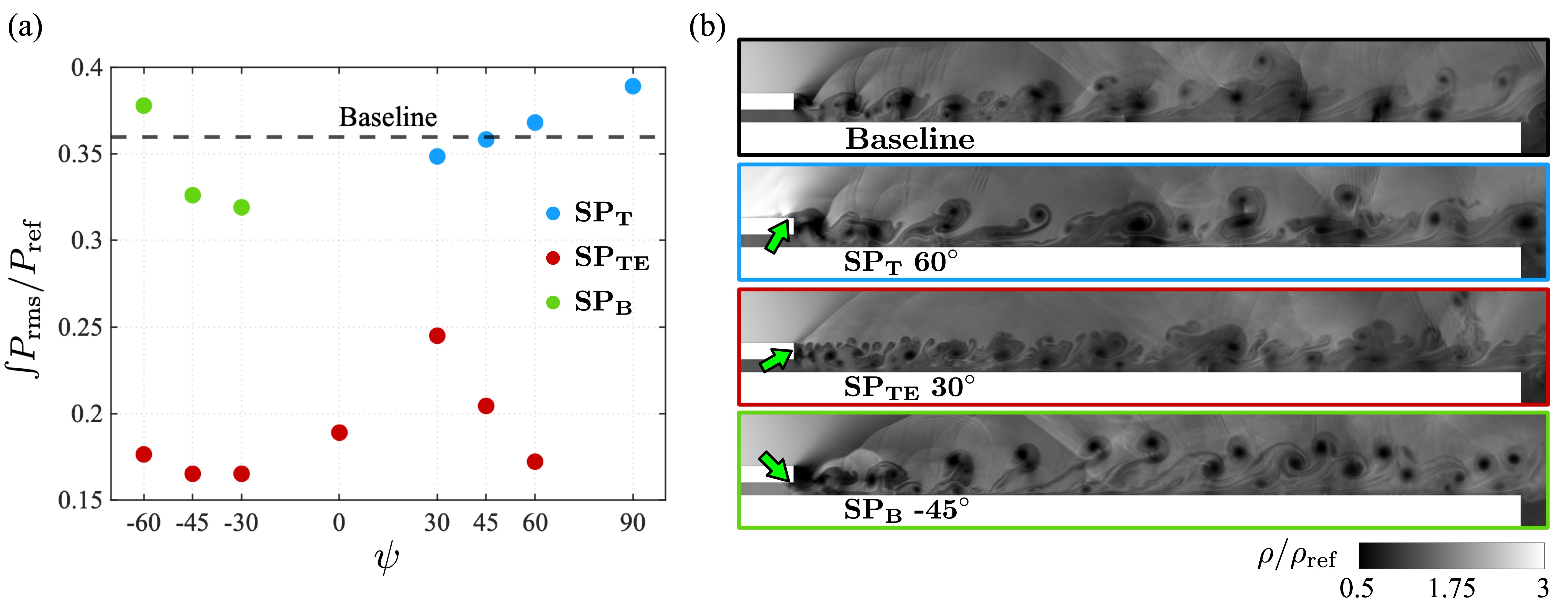}
    \caption{(a) Integrated root-mean-square of pressure fluctuations along aft-deck for all control cases (b) Instantaneous density field for representative control cases. Green arrows indicate micro-jet.}
    \label{fig:Prms_SP}
\end{figure}

To assess the severity of the surface loading along the aft-deck, the integrated pressure fluctuation along the surface from the splitter plate trailing edge to the end of the aft-deck, reported as the root-mean-square after subtracting the mean, is plotted in figure~\ref{fig:Prms_SP}(a) for all cases. Figure~\ref{fig:Prms_SP}(b) shows the instantaneous density field of representative control cases to visualize changes in the trajectory and size of the shedding vortices. When steady blowing is introduced at the $\text{SP}_\text{T}$ location, the resulting surface loading along the aft-deck for all cases is within $\pm~9\%$ of the baseline state. At high angles where $\psi = 60\degree$ and $90\degree$, the upward flow deflection caused by the micro-jet produces slightly larger vortices, which increases the loading along the aft-deck. At low angles where $\psi = 30\degree$ and $45\degree$, the size of the vortices remains comparable to the baseline flow, resulting in minimal change in the surface loading. 

When actuation is introduced at the $\text{SP}_\text{TE}$ location, the vortex-induced surface loading along the aft-deck is reduced across all angles considered due to the weakened contact between the vortices and the aft-deck. The greatest reduction in surface loading occurs when $\psi < 0\degree$. In the uncontrolled flow, the mixing of the main and bypass streams near the SPTE region results in the highest local pressure fluctuation on the aft-deck surface close to the SPTE. When micro-jet actuation is introduced at angles $\psi < 0\degree$, the injected flow continuously impinges on the aft-deck near the SPTE region, leading to smaller local pressure fluctuations. Coupled with the smaller scale vortices, this yields the greatest reduction in the surface loading. As $\psi$ increases from $-30\degree$ to $30\degree$, the lower vortices (see figure~\ref{fig:spte_flow1}(c)) that come into contact with the aft-deck become larger in size, leading to the increase in the surface loading. As $\psi$ increases to $60\degree$, the surface loading begins to decrease. This is likely due to the lower vortices developing slightly above the aft-deck with reduced contact. 

When actuation is introduced at the $\text{SP}_\text{B}$ location, the vortex trajectory is deflected upward, and the vortices convect downstream through the core flow. Since the vortices no longer develop along the aft-deck, there is a moderate reduction in the surface loading. However, this reduction is not as significant as introducing actuation at the $\text{SP}_\text{TE}$ location. This is because the size of the vortices remains comparable to the baseline flow, indicating that influencing the vortex size is most pivotal in reducing the aft-deck surface loading.

\section{Conclusion}
\label{sec:concl}
Control of a supersonic rectangular dual-stream jet flow with steady-blowing micro-jets is examined by parametrically varying spatial locations and injection angles in the vicinity of a splitter plate where the shedding instability due to mixing of the main and bypass streams produces a prominent undesirable resonant tone.  In the baseline flow, the shedding vortices generate compression waves that coalesce to form an oblique shock to deflect the main flow. This relatively strong primary shock traverses through the core flow and impinges on the top of the nozzle wall, where the shock-boundary-layer interaction induces flow separation and affects the upper shear layer developing in the plume downstream. The shedding vortical structures additionally cause high surface loading along the aft-deck. 

A combination of Navier-Stokes, resolvent, and stability analysis is used to explore mitigation and alleviation of the undesired baseline flow features. Discounted resolvent analysis on the turbulent mean baseline flow identifies optimal energy amplification and provides physical insight into effective active flow control configurations. The optimal forcing and response modes indicate that the region near the splitter plate trailing edge is most receptive to external perturbations, motivating the introduction of active control at the splitter plate top, trailing edge, and bottom surfaces, respectively. SPOD of representative control cases identify coherent structures in the flow field, while  discounted resolvent analysis, repeated on all control cases, helps assess forcing-response dynamics changes due to actuation.

When the actuator is placed on the splitter plate top surface, the optimal injection angle $\psi$, measured from the horizontal, to attenuate the shock structures and dominant tone is observed to be $\psi = 60\degree$.  The success of this angle  is traced to the introduction of a new instability upstream of the actuator, detected with SPOD as well as instability analysis of the controlled flow, which predicts a new eigenmode of high growth rate.  The consequent micro-jet-induced waves prevent the main shedding vortex-induced compression waves from coalescing, and the primary shock then no longer forms.  This has consequences on the far nozzle wall, where shock-induced flow separation is muted, and, for all injection angles, the resolvent response modes highlighting the upper shear layer originate closer to the nozzle lip. This indicates that the upper shear layer response is spatially associated with the shock-induced flow separation.

When actuation is introduced on the splitter plate trailing edge surface, the micro-jet breaks the original thick splitter plate shear layer into two thinner shear layers, giving rise to upper and lower sets of shedding vortices. The former generates weaker compression waves, reducing the primary shock strength. Further downstream from the splitter plate trailing edge, the upper and lower sets of vortices interact and merge, forming a secondary shock. Depending on the actuation angle $\psi$, the flow separation near the nozzle lip is induced by either the primary or secondary shock; however, the $\psi = 30\degree$ case achieves a balance between the strength of both shocks, such that neither induces separation. The SPOD analysis also captures the two sets of shedding vortices, which occur at high frequencies relative to the resonant tone found in the baseline flow. Meanwhile, the leading eigenvalue with the lowest growth rate across the baseline and all control cases also occurs at the optimal angle $\psi = 30\degree$, indicating that reducing the size of the shedding vortices and achieving a balance between the primary and secondary shocks can most effectively suppress the baseline instabilities and stabilize the controlled flow. The resolvent modes obtained from all control cases show that the upper shear layer response becomes more unstable and is not sensitive to changes in the angle when $\psi<0\degree$. When $\psi > 0\degree$, the upper shear layer becomes less responsive to perturbations, implying the flow is more stable, however, the upper shear layer is significantly more sensitive to changes in $\psi$. Introducing control with a larger stream-wise velocity component is found to amplify the shock and shear layer instabilities. Actuator placement on the bottom surface of the splitter plate constricts the bypass stream and is not as effective as it amplifies the resonant tone and strengthens shock structures. Similar amplifications in the shock strength are observed for the trailing edge surface actuation in cases where $\psi < 0\degree$. This suggests that to achieve success in control efforts, micro-jet actuation should be introduced upwards into the supersonic main core flow.

The reduction in flow separation, shock strength, and the dominant tone throughout the control cases presented in this study reveals the effectiveness of introducing steady-blowing at the splitter plate proximity region. This prompts future investigations in achieving similar results through the use of unsteady actuators. Applying the discounted resolvent analysis on the controlled flows can provide insights into how the control configuration can be further optimized through a frequency parameter to achieve the same control outcomes with minimal energy expenditure.

\textbf{Acknowledgements.} {We would like to acknowledge the computational resources provided by Syracuse University Research Computing as well as the Syracuse University Fellowship. We also gratefully acknowledge the valuable discussions had with Dr. Mark N. Glauser.}

\textbf{Funding.} {This work is supported by the Air Force Office of Scientific Research under award number FA9550-23-1-0019 (Program Officer: Dr. Gregg Abate).}

\textbf{Declaration of interests.} {The authors report no conflict of interest.}

\textbf{Author ORCIDs.} {M. Yeung, https://orcid.org/0009-0003-3174-3149;}

\bibliographystyle{apalike}  
\bibliography{references}

\end{document}